\documentclass[twocolumn,numberedappendix,iop]{emulateapj}
\shorttitle{Modeling Collisional Cascades: The Numerical Method}
\shortauthors{G\'asp\'ar et al.}
\journalinfo{The Astrophysical Journal, 749:14 (22pp), 2012 April 10}
\usepackage{amsmath}

\begin{document}

\title{Modeling Collisional Cascades In Debris Disks: The Numerical Method}

\author{Andr\'as G\'asp\'ar}
\author{Dimitrios Psaltis}
\author{Feryal \"Ozel}
\author{George H.\ Rieke}
\affil{Steward Observatory, University of Arizona, Tucson, AZ 85721\\
agaspar@as.arizona.edu, dpsaltis@as.arizona.edu, fozel@as.arizona.edu, grieke@as.arizona.edu}

\and

\author{Alan Cooney\altaffilmark{1}}
\affil{Department of Physics, University of Arizona, Tucson, AZ 85721\\
acooney@physics.arizona.edu}

\altaffiltext{1}{also at Steward Observatory, University of Arizona, Tucson, AZ 85721}

\begin{abstract}
We develop a new numerical algorithm to model collisional cascades in
debris disks. Because of the large dynamical range in particle masses,
we solve the integro-differential equations describing erosive and
catastrophic collisions in a particle-in-a-box approach, while
treating the orbital dynamics of the particles in an approximate
fashion. We employ a new scheme for describing erosive (cratering)
collisions that yields a continuous set of outcomes as a function of
colliding masses. We demonstrate the stability and convergence
characteristics of our algorithm and compare it with other treatments. 
We show that incorporating the effects of erosive collisions results 
in a decay of the particle distribution that is significantly faster 
than with purely catastrophic collisions.
\end{abstract}

\section{Introduction}

More than 700 extrasolar planets have been identified to date in over 500 planetary 
systems.\footnote{http://exoplanet.eu} Most of these planets were discovered via radial 
velocity measurements. As a result, only a handful of them are less than 10 Earth masses;
the vast majority are gas giants resembling Jupiter. They are also in extremely close 
orbits to their host stars, making these systems dramatically different from ours.
A large number of additional candidate transiting systems have been found recently with the Kepler 
mission \citep{borucki10}.

In contrast to the great majority of known exoplanet systems, our own solar system has 
a complex configuration with gas giants at significant distances from their central star and
rocky planets/asteroids within the giant planet zone.
The direct detection of rocky planets and planetesimals around other stars is only feasible under very rare circumstances.
One of the most productive approaches is indirectly via the thermal emission of their planetary debris dust belts. 
Ever since the discoveries with IRAS \citep{aumann84,backman93}, we know that extrasolar systems may harbor disks of dust/debris
that are generated by planetesimal collisions. The dust reprocesses the stellar light and emits it as 
thermal radiation in the 10-1000 $\micron$ wavelength 
range. A prototypical example of such a system is Fomalhaut, where a planet is shepherding the 
star's debris disk resolved in both scattered light \citep{kalas08} and in infrared emission 
\citep{holland03,stapelfeldt04,marsh05}. Debris disks highlight the constituents of planetary systems that are 
many to hundreds of AU away from their stars.

With the launch of the {\it Spitzer Space Telescope}, many observations have been obtained
to detect and possibly to resolve debris disks in the infrared regime.
Debris disks have been probed around all types of stars, both in stellar
clusters and in the field. These observations showed that even though debris disks are common around
stars of all spectral types, they are more likely to be detected in the earlier stages of stellar
evolution \citep{wyatt08}. We have also learned that debris disks may be located close to or far from their central
stars, that there are systems with multiple debris rings (such as our solar system), and that there 
can be wide varieties of mineralogical compositions within the disks \citep{carpenter09}. Debris disk studies are now a major
component of the Herschel observing program \citep{matthews10,eiroa10}, which will provide substantial advances in 
our understanding of their outer zones. 

Interpreting these results demands theoretical insights in a variety of areas. For example, attempts 
have been made to understand the evolution of debris disks as a function of stellar type
by studying them in stellar clusters of different ages. As concluded in \cite{gaspar09}, solar-type stars in 
the field \citep{beichman06,trilling08,carpenter08} and in clusters 
\citep{gorlova06,gorlova07,siegler07} may show a faster decay trend compared to that 
observed for earlier-type stars \citep{rieke05,su06}, although the difference is subtle and needs
confirmation. The decay trends of the fractional luminosity 
($f_d = L_{\rm exc}/L_{\ast}$) show a large range in values. \cite{spangler01} find a decay 
$\propto t^{-1.78}$ when fitting ISO/IRAS data, while \cite{greaves03} get a much shallower 
decay $\propto t^{-0.5}$. The majority of surveys however find a decay 
$\propto t^{-1}$ \citep{liu04,moor06,rieke05}. A better theoretical understanding is needed
to sort out these results and to provide testable hypotheses that can be compared with the observations.

Only a handful of debris disks have been resolved; for the majority, we only know the integrated 
infrared excess emission. Finding the underlying spatial distribution
of the debris in these disks is not straightforward, as any spectral energy distribution (SED)
can be modeled with a degenerate set of debris rings at different distances.
Although much of the uncertainty is associated with the optical constants of the grains,
another under-appreciated issue is the grain size distribution. Collisional models can 
reduce the number of free parameters in the SED models by determining
the stable size distribution of particles in the disks.

Observations of resolved debris disks also have raised questions that
can best be addressed by theoretical models. For example, Spitzer MIPS images have shown a significant
extended halo of dust around Vega \citep{su05}, both at 24 and 70 $\micron$.
Initial calculations hypothesized the halo around Vega to be a result of a high
outflow of dust due to radiation pressure from a recent high-mass collisional event \citep{su05}, 
while \cite{muller10} model it as a result of weakly bound particles on highly eccentric orbits.
Further modeling and deep observations of additional systems will help distinguish these two possibilities.

In this paper, we describe a new algorithm for modeling debris disks, in which we refine the physics and 
numerical methods used in collisional cascade models.
In \S 2, we briefly outline previous models and introduce the basics of our algorithm. 
In \S 3 we detail our numerical methods, followed in \S 4 by our approach for
including simplified dynamics.
In the last section we compare our numerical algorithm to previous ones
and discuss in detail the differences between the codes and the effects those
differences have on the outcome of the collisional cascades. We also supplement
our paper with an extended appendix that covers the numerical methods and the
verification tests of our code.

\section{The physical and numerical challenges of modeling debris disks}

Collisional cascades have been studied both analytically and 
using collisional integro-differential numerical models. The classic
analytic models of \cite{dohnanyi69}, \cite{hellyer70}, and \cite{bandermann72}  
took into account both erosive and catastrophic collisional outcomes, assumed a 
material strength that was independent of the particle mass, a particle mass distribution
with no cut-offs, and a constant interaction velocity. These models yielded steady state power-law 
mass distribution indices of -11/6. This result was in general agreement with the 
measured size distribution of asteroids in the solar system. More recently, analytic models by 
\cite{dominik03} and \cite{wyatt07} showed that the fractional infrared luminosity in 
a collision-dominated steady-state system decays following a $t^{-1}$ power-law. As introduced later in this section, \cite{lohne08} find a different
value for this decay timescale. \cite{wyatt07} also derived a maximum mass and fractional luminosity 
as a function of age and distance from the central star, which they then used to classify 
systems with possible recent transient events.

However, numerical models are needed to expand on these results.
In the particular case of our Solar System, sophisticated numerical models were developed
to track the evolution of the largest asteroids \citep{greenberg78}. They have been further improved
to reproduce the observed wavy structure in the size distribution at the very highest 
masses \citep[e.g.,][]{obrien05,bottke05}. These models yield power-law distributions that 
deviate from the classic solution of \cite{dohnanyi69}, with certain regions steeper than 
it and others shallower. Using a steeper or shallower distribution and extrapolating 
it to dust sizes can result in substantial offsets in the number of particles and thus in the infrared emission 
originating from them for a given planetesimal mass. Conversely, the particle size distribution affects
the underlying disk mass calculated from the observed infrared emission.

A complete numerical model of collisional cascades would follow outcomes from all types of collisions,
include a kinematic description of the system, incorporate coagulation below certain
thresholds, and do all this with high numerical fidelity. Although such a model has not yet been built
because of its complexity, there are a number of approaches in the literature 
that model collisional cascades down to particles of micron size, each with distinctive
strengths and weaknesses.

The collisional code {\tt ACE} has been used in many studies \citep{krivov00,krivov05,krivov06,krivov08,lohne08,muller10}.
It follows the evolution of the particle size distributions as well as the spatial distribution 
of the dust in debris disks. The code initially only accounted for collisions resulting in catastrophic 
outcomes, while the latest version \citep{krivov08,muller10} includes erosive (cratering) events as well. 
The collisional outcome prescriptions are based on the \cite{dohnanyi69} particle-in-a-box 
model, but with a more elaborate description of material strengths in collision outcomes as well as the 
radiation force blowout. A strength of the code is that it calculates the dynamical 
evolution of the systems, as well. Since following the dynamical evolution of a system makes large demands on computer memory 
space and CPU speed, the code can only model the size distribution with a low number of mass grid points; it
originally used a first order Euler Ordinary Differential Equation (ODE) solving algorithm, but has been
modified to include a more precise one (A.\ Krivov, priv.\ comm.).
\cite{krivov06} and \cite{muller10} applied this algorithm to debris disks in general and to the specific 
example of the Vega system. They followed the orbital paths of fragments and placed special 
emphasis on radiation effects. \cite{lohne08} modeled debris disk evolution around solar-type 
stars and found, both with analytic and numerical analysis, that the majority of physical 
quantities, such as the mass and the infrared luminosity,
decrease with time as $t^{-0.3}$ to $t^{-0.4}$. This is in contrast with the observed $t^{-1}$ decay
found by some obervations \citep{liu04,moor06,rieke05}. However, the population synthesis verification tests in \cite{lohne08} yield 
good agreement with the latest Spitzer observations.

Th{\'e}bault et al.\ (\citeyear{thebault03}, \citeyear{thebault07}) study 
the evolution of extended debris disks with a particle-in-a-box algorithm. They include both catastrophic 
and erosive collisions and employ resolution and numerical methods similar to the ones implemented in the {\tt ACE} code.
They model the extended disk structure by dividing the disk into separate, but interacting rings. 
Their model does not include dynamics.

\begin{figure*}[!t]
\begin{center}
\includegraphics[angle=0,scale=0.96]{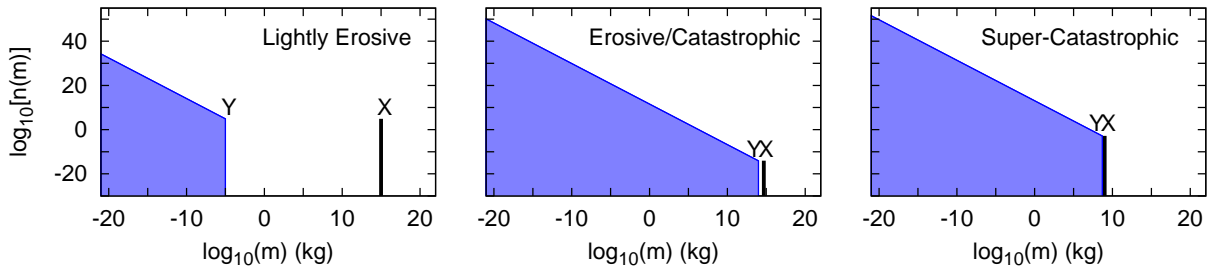}
\caption{Illustration of the possible outcome scenarios of collisions. In all collisions, a single largest $X$ fragment is 
created as well as a power-law distribution of fragments with a largest mass of $Y$.}
\label{fig:toy}
\end{center}
\end{figure*}

\cite{campo94} showed that a series of wave patterns is produced in the mass distribution of particles
when a low-mass cutoff is enforced, such as in the case of a radiation force blowout limit.
This signature is produced by the other debris disk numerical models as well
\citep{thebault03,thebault07,krivov06,lohne08,wyatt11}. However, the conditions under which 
these waves are produced have not been completely analyzed. \cite{wyatt11} do show
that the amplitude and wavelength of the waves is collisional velocity dependent. Such strong 
features in the particle size distribution are not observed in the dust collected within the solar system. The interplanetary
dust flux model of \cite{grun85}, which used in situ satellite measurements of the micro-meteroid
flux in the solar system, and the terrestrial particle flux measurements of the LDEF
satellite \citep{love93} only show a single peak at $\sim 100 \micron$ in the dust distribution.
However, these measurements detected particles that were brought inward from the outer parts of the solar
system via Poynting-Robertson drag and particles removed from the inner parts of the solar system via radiation
force blowout. Their results are reflections of more than a single parent distribution and 
of multiple physical effects.

The dynamical code of \cite{kuchner10} models the evolution and 3D structure of the Kuiper 
belt, with a Monte Carlo algorithm and a simple treatment of particle
collisions. Their models predict that grain-grain collisions are important even
in a low density debris ring such as our Kuiper belt.

\cite{kuchner10} and \cite{dullemond05} both emphasize the strong effects of
fragmentation in their models. \cite{muller10} point out that including
erosive (cratering) events is necessary for their models to reproduce
the observed surface brightness profiles of Vega. \cite{thebault07} also show
that a complete collisional treatment will result in significant deviations from the
classic power-law solution. 

Our goal is to set up a numerical model that places special
emphasis on investigating these issues. Our new empirical description of collisional
outcomes avoids discontinuities between erosive and catastrophic collisions and thus
enables a more stable and accurate calculation. We also solve the full scattering integral, 
thus ensuring mass conservation and the propagation of the largest remnants
of collision outcomes. Finally, we use second order integration and fourth order 
ODE solving methods to improve the numerical accuracy. Below, we outline the 
physical and numerical techniques we will employ. In the Appendix, we present 
verification tests of these treatments.

\subsection{Collisional outcomes}
\label{sec:outcomes1}

In collision theory, two types of outcomes are generally distinguished: catastrophic 
and erosive (the latter also known as cratering). For an erosive collision (EC), the 
collisional energy is relatively small, resulting in one big fragment whose mass is close to 
the original target mass. In the case of a catastrophic collision (CC), 
both colliding bodies are completely destroyed.

We illustrate these outcomes in Figure \ref{fig:toy}.
In the first panel, we plot the outcome distribution of an erosive 
collision, where there is a single large $X$ fragment and a distribution of dust at much lower 
masses. The $X$ fragment is over half the mass of the original target mass $M$. The redistributed 
mass is equal to the cratered mass plus the projectile mass. The largest fragment in 
the distribution, $Y$, is arbitrarily set to be 20\% of the cratered mass. In \S \ref{sec:outcomes}
we elaborate on the validity of this arbitrary value.

In the second panel of Figure \ref{fig:toy}, we plot the outcome at the boundary case between 
catastrophic and erosive collisions, where the single largest fragment $X$ is exactly half of 
the original target mass $M$. The redistributed mass is equal to the other half of the target 
mass $M$ plus the projectile mass. The largest fragment in the distribution, $Y$, is 
arbitrarily set to be 20\% of the cratered mass here as well (10\% of the target mass).

Finally, in the third panel of Figure \ref{fig:toy}, we plot the outcome of a super-catastrophic 
collision, where the target and projectile masses are equal. The mass of the single largest 
fragment $X$ is given by the relation of \cite{fujiwara77}. The redistributed mass is equal 
to $M-X$ plus the projectile mass. The largest fragment in the distribution, $Y$, 
is arbitrarily set to be at $0.5 X$. 

In reality, there is no strict boundary between catastrophic and erosive collisions
\citep{holsapple02}. The outcomes between these two extreme scenarios should be continuous. In laboratory 
experiments, however, it is easier to test the extreme outcomes. In our model, we use the 
laboratory experiments to describe the extreme solutions and connect them with simple 
interpolations throughout the parameter space. We revise the currently used models to
include an $X$ fragment for both erosive and catastrophic collisions as a separate new 
gain term. In this treatment, the placement of the $X$ fragments is grid size independent, further improving precision and guaranteeing
the accurate downward propagation of these fragments. We are able to express the loss term in a much simpler form, 
including collisions from both regimes. Previous models only included a full loss term 
for catastrophic collisions and removed fractions of particles for erosive collisions.
 
The slope of the power-law particle redistribution has been studied extensively. 
\cite{dohnanyi69} used a single power-law value from the largest mass to the smallest. Later 
experiments have shown that a double (or even a triple) power-law distribution is a more 
likely outcome \citep[see, e.g.,][]{davis90}. This has led to the widespread use of a 
double power-law for the redistribution in numerous collision models. 
We conducted numerical tests that demonstrated that there is negligible difference in using
a wide range of slopes with a single power-law. The fact that the outcome does not depend on
the bimodality of the redistribution has also been proven by \cite{thebault03}.
Therefore, we have used the simplest method of redistributing the fragmented particles with a single 
power-law slope from the second largest $Y$ fragment downwards, scaled to conserve mass. As a 
nominal value, we will use -11/6 as the redistribution slope. This
is close to the initial value of -1.8 used by \cite{dohnanyi69}.

\subsection{Incorporating the complete redistribution integral}

The classic solution to the collisional evolution of an asteroid system involves solving the \cite{smol16}
integro-differential equation. This was first done by \cite{dohnanyi69}. Because erosive collisions remove
only a small part of the target mass in a collision, \cite{dohnanyi69} expressed the erosive
removal term in a differential form. This is not 
appropriate for our case. Our system has well defined boundaries; thus a continuity equation
cannot be used. The locality of the collisional outcomes in phase space is not certain either.

Therefore, to solve the Smoluchowski equation for the problem at hand, we need to solve
the full scattering integral. This is complicated numerically, as the integrations must extend over the entire
dynamical range of $\sim 40$ orders of magnitude in mass.
 To be able to perform accurate integrations over such a large interval, we need to use a 
large number of grid points and sophisticated numerical methods. To achieve this in a reasonable time,
we chose to drop the radial dependence of the various quantities and to solve the equations under a
``particle-in-a-box'' approximation. With this approach, we lose radial and velocity information but gain accuracy. 

\subsection{The effect of radiation forces}

Poynting-Robertson drag can be an effective form of removing particles from the disk, so we include it in our 
model. However, the strongest and most dominating radiation effect is the removal of particles via 
radial radiation forces. These act on orbital timescales and can remove or place particles
on extremely eccentric orbits. This gives rise to the challenge of incorporating a radial 
dependent removal term into a particle-in-a-box model that does not carry radial information.
In \S 3.1.1.\ and \S 3.1.2, we discuss our approach for incorporating these radiation effects.

Stellar wind drag is an important dust removal effect for late type stars such as in
the case of AU Mic \citep{augereau06,strubbe06}, an M1 spectral-type star. We concentrate on modeling
debris disks in early- and solar-type systems, so we chose to neglect the effects of stellar wind drag.
We do not take into account the Yarkovsky effect either, as it is small in high-density debris disks compared 
to the other radiation effects.

\begin{figure}[!t]
\begin{center}
\includegraphics[angle=0,scale=0.67]{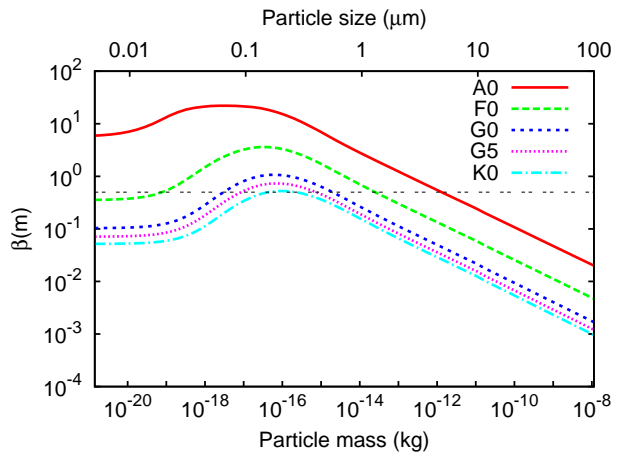}
\caption{Calculated values for the radiation-force parameter $\beta$ around stars of spectral-type A0, F0, G0, G5 and K0. The thin double-dashed 
black line is at the critical value $\beta$=0.5, above which radiation forces are able to remove particles
from circular orbits.}
\label{fig:beta}
\end{center}
\end{figure}

\section{The collisional model}

We now discuss our collisional code ({\tt CODE-M} - {\tt CO}llisional {\tt D}isk {\tt E}volution {\tt M}odel), which
solves the system of 
integro-differential equations that describe the evolution of the number densities of particles of different masses. 
The code includes outcomes from erosive (cratering) collisions and catastrophic collisions and qualitatively follows the 
effects originating from radiation forces and Poynting-Robertson drag.
 
Our system-dependent parameters are: the spectral-type of the central star (which defines the stellar mass $M_{\ast}$
and the magnitude of the radiation effects it will have on the particles), the minimum and maximum particle masses 
($m_{\rm min}$ and $m_{\rm max}$, respectively), the radius, width, and height of the debris ring ($R$, $\Delta R$, 
and $h$, respectively), the total mass within the ring (${\mathcal M}_{\rm TOT}$), and the slope of the initial 
size distribution of the particles ($\eta$). 
We estimate the total volume of the narrow ring, ${\mathcal V}$, as
\begin{equation}
{\mathcal V} = 2 \pi h R \Delta R,
\label{eq:Volume}
\end{equation}
which together with ${\mathcal M}_{\rm TOT}$ defines a mass density.

\subsection{The evolution equation}
\label{sec:evoleq}

In general, the change in the differential number density $n(m,t)$ at any given time for a particle of mass $m$ is given by \citep{smol16}
\begin{equation}
\frac{\rm d}{{\rm d}t} n(m,t) = T_{\rm PRD} + T_{\rm coll}\;,
\label{eq:main}
\end{equation}
where $T_{\rm PRD}$ is the Poynting-Robertson drag (PRD) term and $T_{\rm coll}$ is the sum of the collisional terms.
We define the differential number density of particles such that
\begin{equation}
N(t) = \int n(m,t) {\rm d}m
\end{equation}
is the time-dependent total number density of particles within the ring.

Effects such as radiation force blowout and Poynting-Robertson drag are able to deplete the low-mass 
end of the distribution, which in turn alters the evolution of the disk and more importantly, its infrared signature.
Because we do not follow the radial profile of the various debris disk quantities in our algorithm,
we can only capture the effects of radiation forces in a simplified way.

\subsubsection{Poynting-Robertson drag term}
\label{sec:PRD}

A  complete analysis of the effects of the Poynting-Robertson drag is given by \cite{burns79}, 
who correct many errors made in previous work. 
This effect causes the particles to slow in their orbit 
and follow an inward spiral. \cite{burns79} show that the change in the orbital distance can be written as
\begin{equation}
\frac{{\rm d}R(m)}{{\rm d}t} = -\frac{2 {\rm G} M_{\ast} \beta(m)}{{\rm c} R}\;,
\label{eq:prd}
\end{equation}
where G is the gravitational constant, c is the speed of light, and $\beta(m)$ is the ratio of radiation to gravitational 
force experienced by a particle of mass $m$. 

We calculate the $\beta(m)$ values as a function of the particle masses, optical constants, and the spectral type of the central star 
following \cite{gaspar08}. For the calculations we assume a silicate composition for the particles and a bulk
density of $2.7$ g cm$^{-3}$. We show the calculated $\beta(m)$ values for a few different spectral
type stars in Figure \ref{fig:beta}. 

We use equation (\ref{eq:prd}) to derive an approximate term that captures the effect of Poynting-Robertson drag 
as (see eq.\ [\ref{eq:main}])
\begin{equation}
T_{\rm PRD} = -\frac{n(m,t)}{\tau_{\rm PRD}\left(m\right)}\;,
\end{equation}
where
\begin{equation}
\tau_{\rm PRD}\left(m\right) = \frac{{\rm c}}{2{\rm G} M_{\ast} \beta(m)} R \Delta R\;.
\label{eq:tauprd}
\end{equation}
The mass dependence of the timescale comes from the mass dependence of the parameter $\beta$.
In principle, once a particle is removed from our collisional system it still radiates in the IR; it just does 
not take part in the collisional cascade. We keep track of the removal rate 
of these particles, but do not follow the total amount removed or their infrared emission.

\subsubsection{Radiation force blowout}

The effects of the radiation force blowout are incorporated in our code
with the simplified dynamics treatment introduced in \S 4, and not by the inclusion of a 
separate term in the differential equation as are the effects of 
Poynting-Robertson drag.
Removing a particle from the collisional system via radiation force blowout requires roughly an orbital timescale
\begin{equation}
\tau_{RFB} = 2\pi \sqrt{\frac{R^3}{{\rm G}M_{\ast}}}\;.
\label{eq:taurfb}
\end{equation}
As we will show in \S 4, under our assumptions a newly created particle  of mass $m$ gets 
removed via radiation force blowout if $\beta(m) \ge 0.5$ and is unaffected by radiation forces when $\beta(m) < 0.5$. 

\begin{figure}[!t]
\begin{center}
\includegraphics[angle=0,scale=0.67]{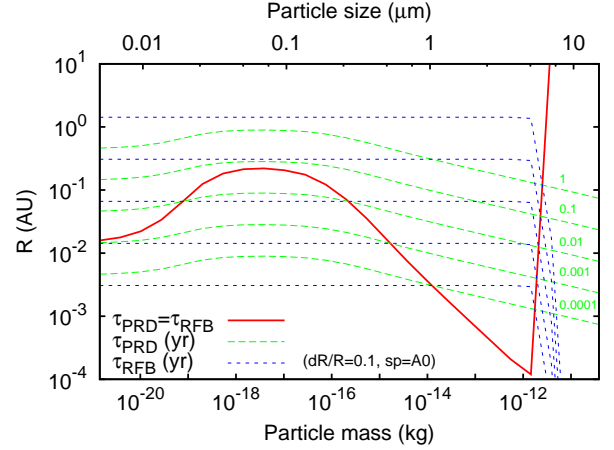}
\caption{Comparison of the radiation force blowout (RFB) to the Poynting-Robertson drag 
(PRD) timescales for an A0 spectral-type star, as a function of particle mass and distance
from the star, with a disk width of d$R/R=0.1$. The dashed lines give the 
orbital distances as a function of particle size, where the Poynting-Robertson drag and blowout
timescales are 0.1, 0.01, 0.001 and 0.0001 years. The solid red line gives the distance where the 
timescale for Poynting-Robertson drag is equal to the radiation force blowout timescale. Above the solid red line, 
radiation force blowout dominates, while below it Poynting-Robertson drag does. 
The plot shows that within reasonable disk radii estimates, radiation force blowout will always 
dominate in the $\beta(m)>0.5$ domain, while outside of it Poynting-Robertson drag will be the stronger effect.}
\label{fig:PRDvsRFB}
\end{center}
\end{figure}

Although the radiation force blowout timescale is not used in our code, in Figure \ref{fig:PRDvsRFB}
we compare it to the Poynting-Robertson drag timescale around an A0 spectral-type star, assuming a disk width-to-radius 
ratio of 0.1. The plot shows that within reasonable disk radii estimates, radiation force blowout will always dominate 
in the $\beta(m)>0.5$ domain, while outside of it Poynting-Robertson drag will be the stronger effect. Whether 
Poynting-Robertson drag is an effective form of removal in the $\beta(m)<0.5$ domain depends on the number density 
of particles in the ring, i.e.,\ the collisional timescale of the system \citep{wyatt05}. The outcomes 
are similar for all spectral type stars and for realistic $\Delta R/R$ values.

\begin{figure*}[!t]
\begin{center}
% ApJ Double Column scaling:
\includegraphics[angle=0,scale=0.7]{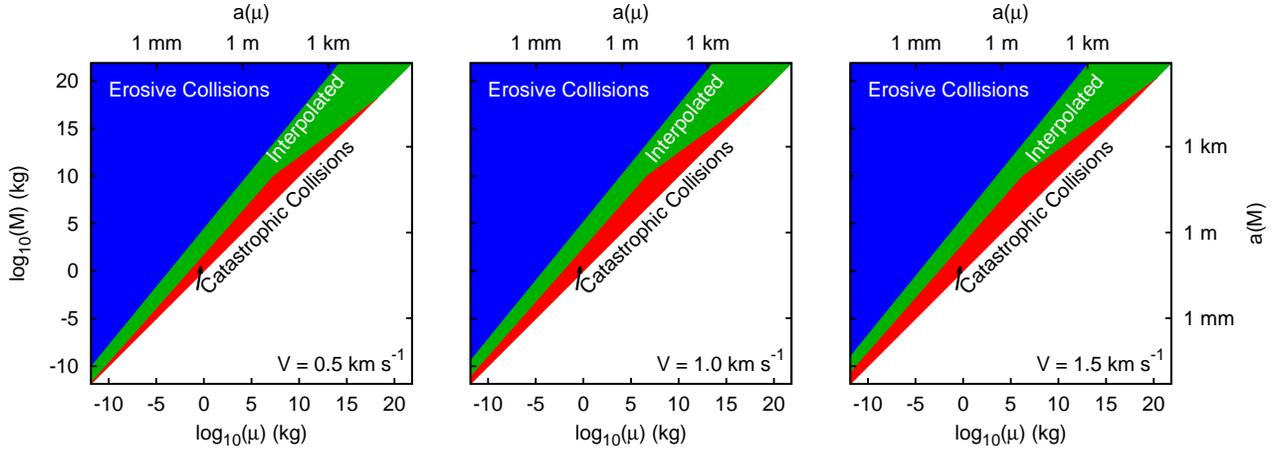}
% ApJ Manuscript scaling:
%\includegraphics[angle=0,scale=0.65]{f4.eps}
\caption{The outcome possibilities as a function of colliding masses plotted for collisional
velocities of 0.5, 1 and 1.5 km s$^{-1}$. These collisional velocities roughly correspond to debris ring 
radii of 100, 25 and 10 AU around an A spectral type star, respectively (see \S 4). We note that higher 
collisional velocities can occur in some systems.}
\label{fig:regions}
\end{center}
\end{figure*}

\subsection{The collisional term}
\label{sec:terms}

The probability of a collision between particles is a function of their number
densities and their collisional cross section. We express the collisional cross section for
particles of mass $m$ and $m^{\prime}$ as
\begin{equation}
\sigma\left(m,m^{\prime}\right) = \pi \left[r\left(m\right)+r\left(m^{\prime}\right)\right]^2\;,
\end{equation}
where $r(m)$ is the radius of each particle. We express the differential rate of collisions between the two masses as
\begin{eqnarray}
P(m,m^{\prime};t) &=& n(m,t)n(m^{\prime},t) V \sigma\left(m,m^{\prime}\right) \nonumber \\
                  &=& n(m,t)n(m^{\prime},t) V \pi \left[r\left(m\right)+r\left(m^{\prime}\right)\right]^2 \nonumber \\
                  &=& \kappa n(m,t)n(m^{\prime},t) V \pi \left(m^{\frac{1}{3}}+m{^{\prime}}^{\frac{1}{3}}\right)^2 
\end{eqnarray}
where $V$ is their characteristic collisional velocity, $n(m,t)$ and $n(m^{\prime},t)$ are the differential number 
densities for particles of mass $m$ and $m^{\prime}$,
\begin{equation}
\kappa \equiv \left(\frac{3}{4\pi\rho}\right)^{\frac{2}{3}}\;,
\label{eq:rho}
\end{equation}
and $\rho$ is the bulk mass density of the particles. The number densities in the problem are naturally 
all time dependent. However, for brevity, hereafter we drop the time dependence from our 
notation.

The decrease or increase in number density at a certain mass will be determined by three
separate events: the removal of particles caused by their interaction
with all other particles, the addition of the $X$ particles from the interactions
of other particles (see \S2.1 and Figure \ref{fig:toy}), and the addition of particles
from the redistribution of smaller fragments originating from collisions of other particles.

We express the first event, which describes the removal of particles, as
\begin{equation}
\frac{\rm d}{{\rm d}t} n(m)\Biggl|_{\rm rem}  = -\int\limits_{m_{\rm min}}^{m_{\rm max}} {\rm d}m^{\prime} P(m,m^{\prime})\;.
\end{equation}
We completely remove all particles from all grid points if they take part in a collision, even if they 
are the target objects in erosive collisions.

The second event to be described is the addition of the large $X$ fragments. To this 
end, we need to calculate the mass $M$ that will produce a particle of mass $m=X$ when interacting 
with a particle of mass $m^{\prime}$. We achieve this with a root finding algorithm from the collisional equations 
presented in the following sections and calculate it only once in the beginning of each run. In equation form,
\begin{equation}
\frac{\rm d}{{\rm d}t} n(m)\Biggl|_{m\equiv{\rm X}(m^{\prime},M)}  = \int\limits_{m_{\rm min}}^{\mu_X(m)} {\rm d}m^{\prime} P(M,m^{\prime})\;.
\end{equation}
The lower limit of the integration is the minimum mass in the distribution. We denote the largest mass $m^{\prime}$ 
that can create a particle of mass $m$ as the $X$ fragment as $\mu_X(m)$. Its value can also be calculated via root 
finding algorithms and has to be calculated for each value of $m$ once in the beginning of each run (see Figure \ref{fig:X} in the Appendix).

These first two integrals may catastrophically cancel, meaning that the difference between the two terms may be significantly smaller than
the absolute value of each, causing the former to be artificially set to zero when evaluated numerically. It is therefore useful to combine 
these terms into a single integral in a way that will lessen the probability of catastrophic cancellation:
%%%%%%%%%%%%%%%%%%%%%%%%%%%%%%%%%%%%%%%%%%%%%%%%%%%%
%
% Manuscript (Single Column) format of equation
%
%%%%%%%%%%%%%%%%%%%%%%%%%%%%%%%%%%%%%%%%%%%%%%%%%%%%
%\begin{eqnarray}
%T_{\rm I}(m)  = && -V \pi \kappa \Biggl\{ \int\limits_{m_{\rm min}}^{\mu_X(m)} {\rm d}m^{\prime}
%n(m^{\prime})\biggl[n(m)\Bigl(m^{\frac{1}{3}}+m{^{\prime}}^{\frac{1}{3}}\Bigr)^2 - n(M)
%\Bigl(M^{\frac{1}{3}}+m{^{\prime}}^{\frac{1}{3}}\Bigr)^2\biggr]\Biggr. \nonumber \\
%                                  && + \Biggl. \int\limits_{\mu_X(m)}^{m_{\rm max}} {\rm d}m^{\prime}
%                                  n(m^{\prime})n(m)\Bigl(m^{\frac{1}{3}}+m{^{\prime}}^{\frac{1}{3}}\Bigr)^2\Biggr\}\;,
%\label{eq:t1}
%\end{eqnarray}
%%%%%%%%%%%%%%%%%%%%%%%%%%%%%%%%%%%%%%%%%%%%%%%%%%%%
%
% ApJ (Double Column) format of equation
%
%%%%%%%%%%%%%%%%%%%%%%%%%%%%%%%%%%%%%%%%%%%%%%%%%%
\begin{eqnarray}
&& T_{\rm I}(m)  = -V \pi \kappa \Biggl\{ \int\limits_{m_{\rm min}}^{\mu_X(m)} {\rm d}m^{\prime}n(m^{\prime})\biggl[n(m)\biggr .\Biggr.\nonumber \\
&& \biggl.\Bigl(m^{\frac{1}{3}}+m{^{\prime}}^{\frac{1}{3}}\Bigr)^2 - n(M)\Bigl(M^{\frac{1}{3}}+m{^{\prime}}^{\frac{1}{3}}\Bigr)^2\biggr] \nonumber \\
&& + \Biggl. \int\limits_{\mu_X(m)}^{m_{\rm max}} {\rm d}m^{\prime} n(m^{\prime})n(m)\Bigl(m^{\frac{1}{3}}+m{^{\prime}}^{\frac{1}{3}}\Bigr)^2\Biggr\}\;,
\label{eq:t1}
\end{eqnarray}
Unfortunately $T_{\rm I}$ can still suffer from catastrophic cancellation (when
$m^{\prime}$ is much smaller than $m$, and by definition only for the first integral). We overcome this issue by
employing a Taylor-series expanded form of $T_{\rm I}$, as given in the Appendix.

The third event in the collisional term is the addition of the power-law fragments back to the distribution. 
The description of this process is quite simple; however, its precise calculation is not. We write in general
%%%%%%%%%%%%%%%%%%%%%%%%%%%%%%%%%%%%%%%%%%%%%%%%%%%%
%
% Manuscript (Single Column) format of equation
%
%%%%%%%%%%%%%%%%%%%%%%%%%%%%%%%%%%%%%%%%%%%%%%%%%%%%
%\begin{equation}
%T_{\rm II}(m) = \int_{m_{\rm min}}^{m_{\rm max}} {\rm d}\mu \int_{\mu}^{m_{\rm max}} {\rm d}
%M P(\mu,M) A(\mu,M)\times H\left[Y(\mu,M)-m\right]m^{-\gamma}\;,
%\label{eq:t2}
%\end{equation}
%%%%%%%%%%%%%%%%%%%%%%%%%%%%%%%%%%%%%%%%%%%%%%%%%%%%
%
% ApJ (Double Column) format of equation
%
%%%%%%%%%%%%%%%%%%%%%%%%%%%%%%%%%%%%%%%%%%%%%%%%%%
\begin{eqnarray}
T_{\rm II}(m) = && \int_{m_{\rm min}}^{m_{\rm max}} {\rm d}\mu \int_{\mu}^{m_{\rm max}} {\rm d}M P(\mu,M) A(\mu,M)\times\nonumber \\
		&& H\left[Y(\mu,M)-m\right]m^{-\gamma} \;,
\label{eq:t2}
\end{eqnarray}
where $\mu$ is the projectile mass, $M$ is the target mass, $A$ is the scaling of the power-law distribution, and $H$ is the Heaviside
function.
The total redistributed mass is
\begin{equation}
M_{\rm redist.}(\mu,M) = \int_0^{Y\left(\mu,M\right)} {\rm A}(\mu,M) m^{-\gamma+1} {\rm d}m\;,
\label{eq:mred}
\end{equation}
where $Y$ is the largest fragment within the redistribution (i.e., the second largest fragment in the collision, after X; see \S2.1). This 
gives the scaling factor
\begin{equation}
{\rm A}(\mu,M) = \frac{\left(2-\gamma\right)M_{\rm redist.}(\mu,M)}{Y^{2-\gamma}\left(\mu,M\right)}\;.
\label{eq:gamma}
\end{equation}
The precision of this integration depends strongly on the resolution of our grid points, due
to the integration limits set by the Heaviside function. We discuss in detail the integration methods we 
used in the Appendix.

\subsection{Collision outcomes}
\label{sec:outcomes}

The collisional equations can be integrated if the values of $X(\mu,M)$, $Y(\mu,M)$, 
and $M_{\rm redist}(\mu,M)$ are known as a function of the colliding masses. Their 
values are strongly dependent on the outcome of the collision they originate from,
which is determined by the energies of the colliding parent bodies. We show the domains
of erosive, interpolated erosive (explained later in the section), and catastrophic 
collisions as a function of the colliding body masses in Figure \ref{fig:regions} 
for collisional velocities of $V=0.5$, 1.0 and 1.5 km s$^{-1}$. 
We introduce the method for calculating collisional velocities from orbital 
velocities in \S 4, but it is a good general approximation that the collisional 
velocity is roughly an order of magnitude smaller than the orbital velocity. These 
collisional velocities will then correspond to debris ring radii of 100, 25 
and 10 AU around an A spectral type star, respectively.

A collision is considered to be catastrophic if
\begin{equation}
Q(\mu,M)_{\rm impact} \equiv \frac{\mu {\rm V}^2}{2 M} \ge Q^{\ast}(M)\;,
\end{equation}
where $Q^{\ast}(M)$ is the dispersion strength parameter of the target mass $M$, $\mu$ 
is the projectile mass, and V is the relative velocity of the projectile compared to the 
parent ring (\S3.1). We use the dispersion strength description of \cite{benz99} and
discuss our choice in the Appendix. Note that, in a more accurate treatment, we would 
redistribute the relative kinetic energy to both masses and not just to the target 
mass (i.e., divide by $\mu+M$ instead of $M$). We are, however, using the original 
definition of $Q_{\rm impact}$ (as opposed to using the relative kinetic energy) because
the $Q^{\ast}(M)$ values that we will be comparing it to were defined the same way 
\citep{benz99} and this definition makes the problem more tractable numerically.
We note that some work has indicated that the tensile strength curve itself may be collision velocity 
dependent \citep{benz99,holsapple02,stewart09}, 
which we currently do not take into account.

In catastrophic collisions both particles are completely destroyed. Based on experimental 
evidence \citep{fujiwara77,matsui84,takagi84,holsapple02}, we will assume that apart from 
the largest fragment $X(\mu,M)$, the total mass is redistributed as a power-law distribution 
of particles up to a mass that we denote as $Y(\mu,M)$. We calculate the largest single mass 
created using the relation \citep{fujiwara77}
\begin{equation}
X(\mu,M) = M \frac{1}{2}\left[\frac{\mu {{\rm V}}^2}{2 M Q^{\ast}(M)}\right]^{-\beta_{\rm X}}\;.
\label{eq:fujiwara}
\end{equation}
At the separatrix between catastrophic and erosive collisions, $Q(\mu,M)_{\rm impact} = Q_{\rm D}^{\ast}(M)$, 
and $X(\mu,M)=M/2$, which is exactly what we expect. The $\beta_{\rm X}$ factor is measured 
to be 1.24 by \cite{fujiwara77} and this is the fiducial value that we use. Some experiments 
have shown that the shape and material of the target have an effect on the exact value of 
$\beta_{\rm X}$ \citep{matsui84, takagi84}. We will elaborate on the effects of 
varying $\beta_{\rm X}$ in an upcoming paper.

The second largest fragment, $Y$, is always a fraction ($0 < f_Y < 1.0$) of the 
cratered mass, $M_{\rm cr}$, in the erosive collision domain up to the erosive/catastrophic 
collision boundary. In the catastrophic collision domain, $Y$ is a fraction ($f_X$) of the 
$X$ fragment. We interpolate $f_X$ from its value defined by $f_Y$ at the separatrix 
(where $f_Y=f_X$, as $X=M_{\rm cr}$) to a specified value $f_X^{\rm max}$ at the super 
catastrophic collision case of $\mu=M$ as
\begin{equation}
f_X = \exp\left\{\ln\left(f_Y\right) + \ln\left(\frac{f_X^{\rm max}}{f_Y}\right)\frac{\ln\left[\frac{\mu}{2Q^{\ast}(M)MV^{-2}}\right]}{\ln\left[\frac{M}{2Q^{\ast}(M)MV^{-2}}\right]}\right\}\;.
\label{eq:interpol2}
\end{equation}
Our fiducial values for these fractions are $f_Y=0.2$ and $f_X^{\rm max}=0.5$. We express 
the remaining mass in catastrophic collisions as
\begin{equation}
M_{\rm redist}(\mu,M) = \mu + M - X(\mu,M)\;,
\end{equation}
which is redistributed as a large number of smaller particles.

Erosive collisions are more complicated and less well understood. A collision will be erosive if
\begin{equation}
Q(\mu,M)_{\rm impact} \equiv \frac{\mu {\rm V}^2}{2 M} < Q_{\rm D}^{\ast}(M)\;.
\end{equation}
As described in \S 3, an erosive collision will result in a single large fragment, which will be
a remnant of the target body, and a distribution of smaller particles. We use the formula of 
\cite{koschny01a,koschny01b}, i.e.,
\begin{equation}
M_{\rm cr} = \alpha \left[\frac{\mu V^2}{2}\right]^{b}\;,
\label{eq:koschny}
\end{equation} 
to calculate the total mass cratered from the target, where $\alpha$ and $b$ are constants, with fiducial 
values of $\alpha=2.7\times10^{-6}$ and $b=1.23$. This formula is only valid for small cratered 
masses; it can lead to artificially high values for the cratered masses 
(much larger than the target mass) even in
the erosive collision domain. When the cratered mass given by this formula is larger than an arbitrarily set 
fraction $f_M$ of the target mass, we use the following interpolation formula
\begin{equation}
M_{\rm cr}=M\times\exp\left\{\ln(f_M)+\ln\left(\frac{0.5}{f_M}\right)\frac{\ln\left(\frac{\mu V^2}{2M}/Q_l\right)}{\ln\left[Q_{\rm D}^{\ast}(M)/Q_l\right]}\right\}\;,
\label{eq:interpol1}
\end{equation}
where
\begin{equation}
Q_l=\left(\frac{f_M}{\alpha} M^{1-b}\right)^{1/b}\;.
\label{eq:Ql}
\end{equation}
We choose an arbitrary fiducial value for $f_M$ of $10^{-4}$. 

\begin{deluxetable*}{llrl}
\tablewidth{0pt}
\tablecolumns{4}
\tablecaption{THE NUMERICAL, COLLISIONAL AND SYSTEM PARAMETERS USED IN OUR MODEL AND THEIR FIDUCIAL VALUES \label{tab:tabvar}}
\tablehead{
\colhead{Variable} & \colhead{Description} & \colhead{Fiducial value} & \colhead{Notes} }
\startdata
\multicolumn{4}{c}{Numerical variable}\\
\hline\hline
$\delta$	 & Neighboring grid point mass ratio \dotfill					& 1.1				& \S C.2 \\
\hline
\multicolumn{4}{c}{System variables} \\
\hline\hline
$\rho$		 & Bulk density of particles \dotfill					& 	& Eq.\ (\ref{eq:rho}) \\
$m_{\rm min}$ 	 & Mass of the smallest particles in the system \dotfill		&  	& \S 3, Eq.\ (\ref{eq:t1}) \\
$m_{\rm max}$ 	 & Mass of the largest particles in the system \dotfill			& 	& \S 3, Eq.\ (\ref{eq:t1}) \\
$M_{\rm tot}$	 & The total mass within the debris ring \dotfill			& 	& \S 3\\
$\eta$		 & Initial power-law distribution of particle masses \dotfill		& 	& \S 3\\
$R$		 & The distance of the debris ring from the star \dotfill 		& 	& Eqs.\ (\ref{eq:Volume}, \ref{eq:tauprd}, \ref{eq:taurfb}, \ref{eq:em}, \ref{eq:w}) \\
$\Delta R$	 & The width of the debris ring \dotfill				& 	& Eqs.\ (\ref{eq:Volume}, \ref{eq:tauprd}, \ref{eq:em}, \ref{eq:w}) \\
$h$		 & The height of the debris ring \dotfill				& 	& Eqs.\ (\ref{eq:Volume}, \ref{eq:iota}) \\
Sp	 	 & The spectral-type of the star \dotfill				& 	& \S 4 \\
\hline
\multicolumn{4}{c}{Collisional variables}\\
\hline\hline
$\gamma$	 & Redistribution power-law \dotfill						& 11/6 				& Eqs.\ (\ref{eq:t2}, \ref{eq:mred}, \ref{eq:gamma}) \\
$\beta_X$	 & Power exponent in X particle equation \dotfill				& 1.24 				& Eqs.\ (\ref{eq:fujiwara}, \ref{eq:XmuM}) \\
$\alpha$	 & Scaling constant in $M_{\rm cr}$ \dotfill					& $2.7\times10^{-6}$ 		& Eqs.\ (\ref{eq:koschny}, \ref{eq:Ql}) \\
$b$		 & Power-law exponent in $M_{\rm cr}$ equation \dotfill				& 1.23 				& Eqs.\ (\ref{eq:koschny}, \ref{eq:Ql}) \\
$f_M$		 & Interpolation boundary for erosive collisions \dotfill			& $10^{-4}$ 			& Eqs.\ (\ref{eq:interpol1}, \ref{eq:Ql}) \\
$f_Y$		 & Fraction of $Y/M_{\rm cr}$ \dotfill						& 0.2 				& Eq.\ (\ref{eq:interpol2}, \ref{eq:Y}, \ref{eq:fYmuM}) \\
$f_X^{\rm max}$  & Largest fraction of $Y/X$ at super catastrophic collision boundary \dotfill 	& 0.5 				& Eq.\ (\ref{eq:interpol2}) \\
$\Theta$	 & Constant in smoothing weight for large-mass collisional probability\dotfill	& $10^6 m_{\rm max}$		& Eq.\ (\ref{eq:sigmaw}) \\
$P$		 & Exponent in smoothing weight for large-mass collisional probability\dotfill	& 16				& Eq.\ (\ref{eq:sigmaw}) \\
$Q_{\rm sc}$	 & The total scaling of the $Q^{\ast}$ strength curve \dotfill			& 1 				& Eq.\ (\ref{eq:BA99}) \\
$S$		 & The scaling of the strength regime of the $Q^{\ast}$ strength curve \dotfill	& $3.5\times10^{7}$ erg/g 	& Eq.\ (\ref{eq:BA99}) \\
$G$		 & The scaling of the gravity regime of the $Q^{\ast}$ strength curve \dotfill	& 0.3 erg cm$^3$/g$^2$		& Eq.\ (\ref{eq:BA99}) \\
$s$		 & The power exponent of the strength regime of the $Q^{\ast}$ strength curve \dotfill & -0.38 			& Eq.\ (\ref{eq:BA99}) \\
$g$		 & The power exponent of the gravity regime of the $Q^{\ast}$ strength curve \dotfill & 1.36 			& Eq.\ (\ref{eq:BA99})
\enddata
\end{deluxetable*}

In erosive collisions, the single large fragment is expressed as
\begin{equation}
X=M-M_{\rm cr}\;.
\end{equation}
As defined before, the largest fragment of the redistributed mass is a fraction $f_Y$ of the cratered mass
\begin{equation}
Y(\mu,M) = f_Y M_{\rm cr}(\mu,M)\;,
\label{eq:Y}
\end{equation}
while the redistributed mass is
\begin{equation}
M_{\rm redist}(\mu,M) = \mu + M_{\rm cr}\;.
\end{equation}
Thus, the $X(\mu,M)$, $Y(\mu,M)$, and $M_{\rm redist}(\mu,M)$ parameters can be summarized as
\begin{eqnarray}
X(\mu,M) &=&
\begin{cases}
M \frac{1}{2}\left[\frac{\mu {{\rm V}}^2}{2 M Q^{\ast}(M)}\right]^{-\beta_{\rm X}} & \mbox{in CC}\\
M - M_{\rm cr}(\mu,M) & \mbox{in EC}
\label{eq:XmuM}
\end{cases}\\
Y(\mu,M) &=& 
\begin{cases}
f_X(\mu,M)X(\mu,M) & \mbox{~~in CC}\\
f_Y M_{\rm cr}(\mu,M) & \mbox{~~in EC}
\label{eq:fYmuM}
\end{cases}\\
M_{\rm redist}(\mu,M) &=& 
\begin{cases}
\mu + M - X(\mu,M) & \mbox{~~in CC}\\
\mu + M_{\rm cr}(\mu,M) & \mbox{~~in EC}
\end{cases}
\end{eqnarray}
The $M_{\rm redist}$ is redistributed as a large number of smaller particles in both collision 
types, with a slope of $\gamma=11/6$ and a scaling given by equation (\ref{eq:gamma}). We choose 
a redistribution slope of $\gamma=11/6$, which is a value close to that given by 
experimental results \citep{davis90} and is the same as used by \cite{dohnanyi69}.

We give a list of the variable collisional parameters of our model and their fiducial values 
in Table \ref{tab:tabvar}.

\subsection{The initial distribution and fiducial parameters}

We use the \cite{dohnanyi69} steady-state solution of $\eta=11/6$ as our initial distribution, where
$\eta$ is the slope of the initial distribution and yields an initial number density of $n(m) = {\rm C} m^{-\eta}$, where
C is an appropriate scaling constant for the distribution. The exact value of this slope
is unknown for all real systems. Fortunately, the convergent solutions and the timescales of reaching a convergent
solution are fairly insensitive to this value. 

\section{Simplified Dynamics}

For the smallest particles, which we are particularly interested in modeling, radiation forces lead to effects such as
reduced collisional probabilities in thin ring disks and increased collisional velocities in extended disks. In this section, we describe our
approximate treatment of these effects.

The radiation originating from the central star effectively modifies the mass of the star seen by the particles; 
the orbits themselves remain conic sections. The effective mass of the star is decreased by a factor of
$1-\beta(m)$, where $\beta(m)$ is defined in \S \ref{sec:PRD}. If
\begin{equation}
\beta(m) \ge \frac{1}{2}
\end{equation}
a newly created particle is generally put on a hyperbolic orbit, which we take as the requirement for 
radiation force blowout to occur \citep[][and references therein]{kresak76,burns79}.

The effects of dynamical evolution on the collisional cascade can be traced to the eccentricity 
of the orbits. To follow the orbital path of a dust grain that has been created in a collision, we 
assume that the parent bodies were on circular orbits at radius $R$ 
and had $\beta(m)\approx0$.  We also assume that the produced grain will be created with very small relative velocity with respect to
the parent bodies. Under these conditions, it can be shown that the semi-major axis of the acquired orbit will be
\begin{equation}
a(m) = \frac{1-\beta(m)}{1-2\beta(m)}R\;.
\end{equation}

\begin{figure*}[!t]
\begin{center}
% ApJ Double Column scaling
\includegraphics[angle=0,scale=0.67]{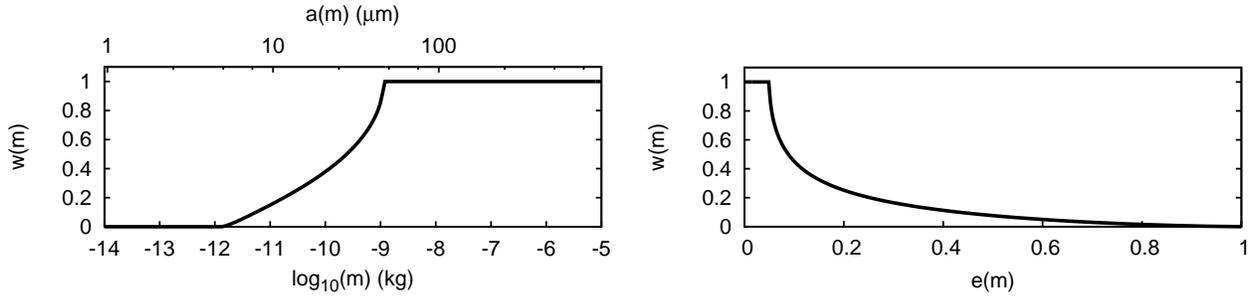}
% ApJ Manuscript scaling
%\includegraphics[angle=0,scale=0.63]{f5.eps}
\caption{{\it Left Panel:} The weighted collisional probabilities of particles as a function of their mass $m$ (and size).
{\it Right Panel:} The weighted collisional probabilities of particles as a function of their eccentricity $e$. The particles
are assumed to be within a narrow (\mbox{${\rm d}R/R=0.1$}) ring at 25 AU from an A0 spectral-type star.}
\label{fig:weights}
\end{center}
\end{figure*}

At $\beta(m) = 0.5$ the semi-major axis becomes infinite, while at
$\beta(m)=0$ it is equal to the semi-major axis of the colliding particles' original orbit. The 
eccentricity ($e_{\beta}$) of the orbit can be determined from the fact that the periapsis 
will equal the original orbital distance
\begin{equation}
a(m)\left[1-e_{\beta}(m)\right] = R\;,
\end{equation}
yielding
\begin{equation}
e_{\beta}(m) =
\begin{cases}
\beta(m)/\left[1-\beta(m)\right] & \mbox{if} ~\beta \le 0.5\\
>1 & \mbox{if} ~ \beta > 0.5
\end{cases}\;.
\end{equation}
At $\beta(m)=0.5$, the eccentricity equals 1, and at $\beta(m)=0$, it equals zero, consistent with our expectations. 
Similar derivations can be found elsewhere \cite[e.g.,][]{harwit63,kresak76}.  A particle on an eccentric orbit 
will have a modified probability of interaction with other particles in the parent ring, which we address in \S 4.2.

\subsection{Collisional velocities}
\label{sec:v}

\cite{lissauer93} give the velocity of a planetesimal relative to the other
planetesimals in the swarm (i.e., the collisional velocity), averaged over an epicycle and over a 
vertical oscillation as
\begin{equation}
V = v_{\rm orb}\sqrt{\frac{5}{4}\left(\frac{e}{2}\right)^2 + \left(\frac{i}{2}\right)^2}\;,
\label{eq:lissauer}
\end{equation}
where $e$ is the maximum eccentricity and $i$ is the maximum inclination in the system.
This equation is valid for a swarm of particles in Rayleigh distributed equilibrium. This condition
is true for a system in quasi-collisional equilibrium. We use this equation to estimate the
collisional velocity of all particles, setting
\begin{equation}
e = \frac{\Delta R}{2R}
\label{eq:em}
\end{equation}
and
\begin{equation}
i = \frac{h}{2R}\;.
\label{eq:iota}
\end{equation}

The smallest particles that are in highly eccentric orbits will have varying velocities along 
their trajectories. However, when at their periapsis, they will have their original orbital velocities, as by
definition they are on eccentric orbits due to their original periapsis velocity. Because of this, in our
simplified dynamical treatment we only use a single collisional velocity for all particles, which
is described by equation (\ref{eq:lissauer}).

\subsection{Reduced collisional probabilities of $\beta$ critical particles}

Particles with $\beta(m)$ less than 0.5, but which are still non-zero, called $\beta$ critical particles,
are thought to produce halos around debris disks via the highly eccentric orbits radiation forces place them 
on \citep{thebault08,muller10}.

For a particle to go into an eccentric orbit, it must acquire a radial velocity component that is different
than zero. In collisions, fragments will be ejected in all directions
with a certain velocity distribution. Since the smallest fragments will tend to escape with the highest 
velocities \citep[e.g.,][]{jutzi10}, it is a fair question to ask whether thermalization of velocity vectors and their
high values is a stronger effect in placing dust particles on higher eccentricity orbits compared to radiation 
effects that reduce the effective stellar mass. 

To answer this question, we need to examine the origin of the particles that contribute to the increase 
of the differential density in each mass grid point. We calculate $T_{II}$ only integrating
in $\mu$ space, thus calculating the rate of increase of the differential number density of particles with
mass $m$ that originate from collisions with targets of mass $M$. 
Our  calculations show that there is a pronounced peak in $M$ roughly on the 
same scale (at most one order of magnitude higher) as $m$ itself. That is, most particles originate from targets $\sim 3-5\times$ 
larger in size than the particle itself. The results of \cite{jutzi10} clearly show that the velocities acquired by collision fragments at 
1/3 sizes are more than an order of magnitude lower than the collisional velocities, meaning that, in the most extreme case, a fragment 
will receive up to a few tenths of a km s$^{-1}$ radial velocity compared to its 10-30 km s$^{-1}$ orbital velocity. 
We can thus safely say that particles that are created with $\beta(m)$ values similar to $0.5$ tend to be placed 
on eccentric orbits by the radiation forces rather than being dispersed. These orbits will extend out from 
the initial debris disk ring, preventing the particles from being destroyed and from them creating other particles. 

Our approach to calculating a weighting factor for each particle mass, determined by the 
fraction of its orbital period it spends in the parent ring is similar to that of \cite{thebault08}. The orbital 
time of a particle in an elliptical orbit as a function of its distance from 
the center of mass is \citep{taff85}
\begin{equation}
\cos^{-1}\left(\frac{a-l}{ae_{\beta}}\right) - e_{\beta}\sqrt{1-\left(\frac{a-l}{ae_{\beta}}\right)^2} = (t-t_0)\sqrt{\frac{GM}{a^3}}\;,
\end{equation}
where $t_0=0$ is the initial time at periapsis, $l$ is its distance at time $t$ from the center of mass, and we omit the 
$m$ dependences of $e_{\beta}$ and $a$ for clarity. 
We estimate the semi-major axis as
\begin{equation}
a = \frac{R-\Delta R/2}{1-e_{\beta}}\;,
\end{equation}
and we calculate the time $\Delta t$ needed for a particle to reach the outer edge of the disk at $l=R + {\rm d}R/2$.
Dividing $\Delta t$ by the half of the orbital period gives the weighting factor for each mass $m$ as
%%%%%%%%%%%%%%%%%%%%%%%%%%%%%%%%%%%%%%%%%%%%%%%%%%%%
%
% Manuscript (Single Column) format of equation
%
%%%%%%%%%%%%%%%%%%%%%%%%%%%%%%%%%%%%%%%%%%%%%%%%%%
%\begin{equation}
%w=\frac{1}{\pi}\left\{\cos^{-1}\left[\frac{\Delta R\left(2-e_{\beta}\right)-2Re_{\beta}}{e_{\beta}\left(\Delta R-2R\right)}\right]-2 \sqrt{\frac{\Delta R\left(e_{\beta}-1\right)\left(\Delta R-2Re_{\beta}\right)}{\left(\Delta R-2R\right)^2}}\right\}
%\label{eq:w}
%\end{equation}
%%%%%%%%%%%%%%%%%%%%%%%%%%%%%%%%%%%%%%%%%%%%%%%%%%%%
%
% ApJ (Double Column) format of equation
%
%%%%%%%%%%%%%%%%%%%%%%%%%%%%%%%%%%%%%%%%%%%%%%%%%
\begin{eqnarray}
w&=&\frac{1}{\pi}\Biggl\{\cos^{-1}\left[\frac{\Delta R\left(2-e_{\beta}\right)-2Re_{\beta}}{e_{\beta}\left(\Delta R-2R\right)}\right] \Biggr.\nonumber \\
&& \Biggl. - 2 \sqrt{\frac{\Delta R\left(e_{\beta}-1\right)\left(\Delta R-2Re_{\beta}\right)}{\left(\Delta R-2R\right)^2}}\Biggr\}
\label{eq:w}
\end{eqnarray}
We plot these weighting factors as a function of the particle mass and orbital eccentricity in Figure \ref{fig:weights}. 
When analyzing the particle distributions, we only plot the number of particles within the parent 
ring, which we calculate as
\begin{equation}
n_{\rm ring}(m) = n(m)w(m)\;.
\end{equation}

\subsection{Reduced collisional probabilities of the largest particles}
\label{largemass}

The very last grid point in the domain of solution will only reduce its number density, as it cannot gain from larger
masses either as an $X$ fragment or from being part of a redistribution. The
full phase space removal, introduced in \S \ref{sec:terms}, causes its evolution time to become very small compared 
to all others and leads to a numerical instability. In order to avoid this, we multiply the collisional
rates with a weight that smooths to zero for the largest particles
\begin{equation}
\sigma_w(m)=\left[\frac{1-{\rm exp}\left(-\frac{m_{\rm max}-m}{\Theta}\right)}{1-{\rm exp}\left(-\frac{m_{\rm max}}{\Theta}\right)}\right]^p\;,
\label{eq:sigmaw}
\end{equation}
for both the projectile and target particle. We chose $\Theta$ to be a number a few orders of magnitude larger than 
$m_{\rm max}$ and use an arbitrary $p=16$. The modified collisional rates, therefore, read
\begin{eqnarray}
P(m,m^{\prime}) &=& V \pi \kappa n(m)n(m^{\prime})w(m)w(m^{\prime}) \times \nonumber \\
		&&\sigma_w(m)\sigma_w(m^{\prime})\left(m^{\frac{1}{3}}+m{^{\prime}}^{\frac{1}{3}}\right)^2\;.
\end{eqnarray}
We discuss the implications of our choice of the weight function and of its parameters in \S 5.2.

\section{Results}

As we discussed in \S 2, collisional cascades in debris disks have been studied extensively in the 
past decades, with many different analytic and numerical solutions to the problem. To demonstrate the
similarities and differences between our model and some earlier ones, we show in the following subsection the
results of a few comparison tests. The system variables used by our code for these runs are summarized in 
Table \ref{tab:comp}. 

We compare our numerical model to three previous well known algorithms, the particle-in-a-box code 
of \cite{thebault03}, the dynamical code {\tt ACE} \citep{krivov00,krivov05,lohne08,muller10}, and
the 1D steady-state solver code of \cite{wyatt11}. Although we do make an effort to model their 
systems as accurately as possible, a true benchmark between the codes is impossible. This is 
due in part to the fact that all models have somewhat different collisional and dynamical prescriptions.

\begin{deluxetable}{lll}
\tablewidth{0pt}
\tablecolumns{3}
\tablecaption{PARAMETERS USED FOR COMPARISON MODELS\label{tab:comp}}
\tablehead{
\colhead{Variable} & \colhead{Comparison to}     & \colhead{Comparison to} \\ 
\colhead{}              & \colhead{Th\'ebault (2003)} & \colhead{L\"ohne (2008)}
}
\startdata
$\rho$ (kg m$^{-3}$)			& 2700 				& 2500			\\
$m_{\rm min}$ (kg)	 	 		& 1.42$\times10^{-21}$ 	& 1.42$\times10^{-21}$	\\
$m_{\rm max}$ (kg)	 	 		& 1.78$\times10^{18}$ 	& 4.20$\times10^{18}$	\\
$M_{\rm tot}$ (M$_{\earth}$)		& 0.0030221 			& 1.0			\\
$\eta$			 			& 11/6				& 1.87			\\
$R$ (AU)			 			& 5 					& 11.25			\\
$\Delta R$ (AU)		 		& 1 					& 7.5			\\
$h$ (AU)			 			& 0.5 				& 3.4			\\
Sp		 	 				& A0					& G5			\\
PRD							& off					& off
\enddata
\end{deluxetable}

\subsection{Comparison to \cite{thebault03}}

A relatively straightforward comparison can be made between {\tt CODE-M} and the \cite{thebault03} model.
Although the initial \cite{thebault03} model has been subsequently improved in \cite{thebault07}, we
chose to compare our results with the former, as they are both particle-in-a-box approaches to the collisional 
cascade, with some dynamical effects included in a simplified manner. 

\cite{thebault03} aimed to model the inner 10 AU region of the $\beta$ Pictoris
disk, with their reference model being a dense debris ring at 5 AU, with a width of 1 AU and height of 0.5 AU.
We adopted their largest particle size of 54 km for the comparison run. However, we adopted 
a smaller minimum particle size than they did (in our case well below the blowout size), to be 
able to follow the removed particles more completely. This choice
does not affect the actual distribution within the ring. We also had
Poynting-Robertson drag turned off. Although both of our models include erosive (cratering) collisions,
the \cite{thebault03} prescription uses hardness constants ($\alpha$) of much softer material than that of our nominal 
case and a linear relationship between cratered mass and impact energy (the prescription for erosive collisions has been changed 
in their later paper Th\'ebault \& Augereau \citeyear{thebault07}). For a better agreement, we also
model a modified cratered prescription case, where we set $b=1$, $\alpha=10^{-4}$ and $f_M=0.01$. With
these adjustments our cratering prescriptions agree better; however our interpolation formula is
offset compared to \cite{thebault03}. While ours has a continuous prescription at the CC/EC boundary
(i.e., the cratered mass is $0.5 M$), the \cite{thebault03} interpolation does not (i.e., the cratered mass
is $0.1 M$). \cite{thebault07} improve on this, by 
employing an interpolation formula very similar to our equation (\ref{eq:interpol1}).

\begin{figure*}[!t]
\begin{center}
% ApJ Double Column scaling
\includegraphics[angle=0,scale=0.75]{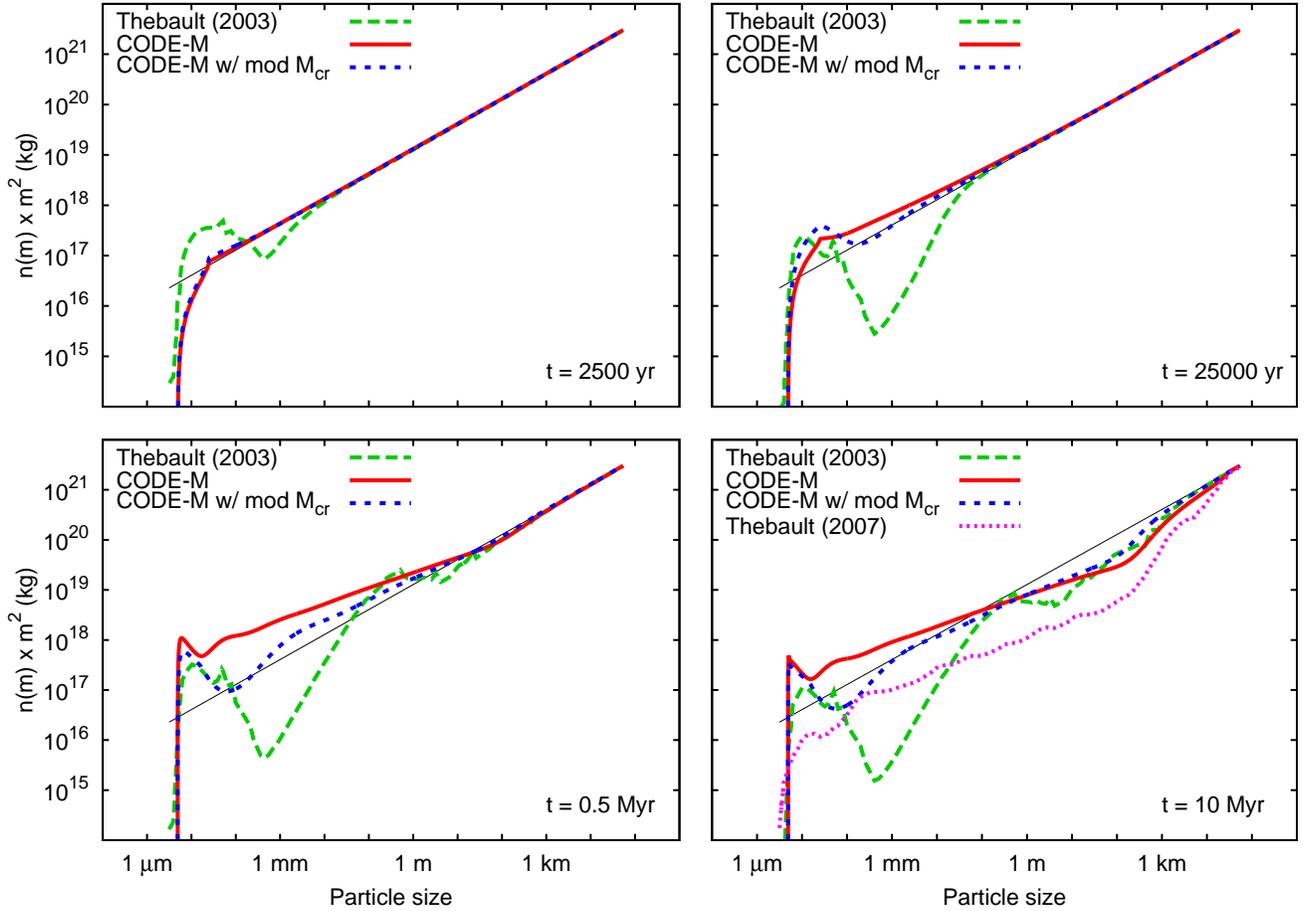}
% ApJ Manuscript scaling
%\includegraphics[angle=0,scale=0.7]{f6.eps}
\caption{Comparison of the evolution of the dust distribution around the $\beta$ Pictoris
disk modeled by \cite{thebault03} and the model presented in this paper. The thin solid line
is the initial distribution. The modified cratered mass ($M_{cr}$) model uses an erosive collision 
prescription that is a closer analog to the original \cite{thebault03} soft material one. We also
plot the curve from the innermost annulus of \cite{thebault07} as reference, scaled to the same
density at the largest sizes as our model. We emphasize though that the \cite{thebault07} 
model is of an extended system, which is significantly different than the \cite{thebault03} model or
ours and is only plotted as a reference. It is noteworthy, however, that the dynamics and extendedness
of the disk structure seemingly have less effect on the results than the material properties or the collisional 
prescription itself.
See text for more details.}
\label{fig:thebaultm}
\end{center}
\end{figure*}

\begin{figure*}[!t]
\begin{center}
% ApJ Double Column scaling
\includegraphics[angle=0,scale=0.7]{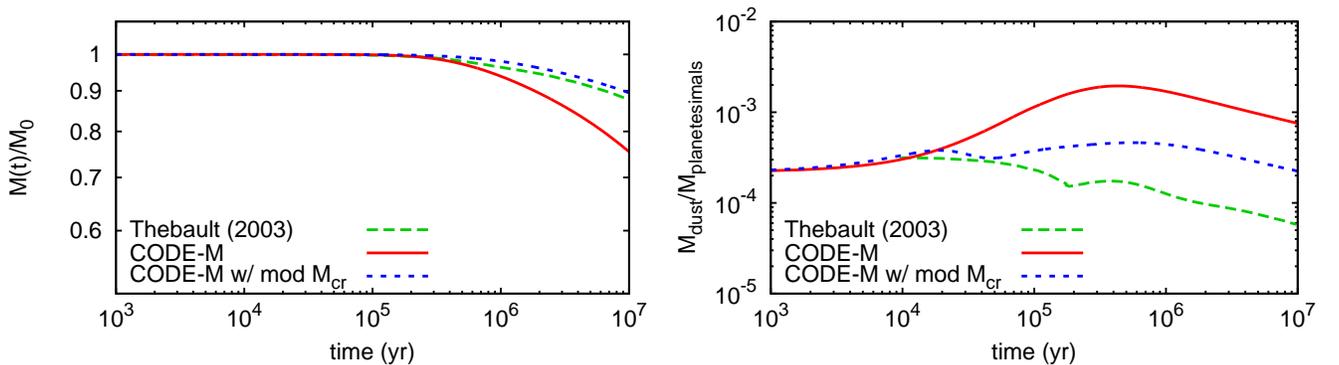}
% ApJ Manuscript scaling
%\includegraphics[angle=0,scale=0.65]{f7.eps}
\caption{Comparison of the evolution of the total disk mass and the dust-to-planetesimal mass ratio around 
the $\beta$ Pictoris disk modeled by \cite{thebault03} and the models presented in this paper. The modified 
cratered mass ($M_{cr}$) model uses an erosive collision prescription that is a closer analog to the original 
\cite{thebault03} one. See text for more details.}
\label{fig:thebaultmass}
\end{center}
\end{figure*}

Figure \ref{fig:thebaultm} compares the evolution of the distribution of particles 
between the \cite{thebault03} nominal case and our runs. In the vertical axes we plot $n(m) \times m^2$, which
is similar to the ``mass/bin" value used by \cite{thebault03}. To make them exact, we divide the \cite{thebault03}
values by ($\delta$-1), which places them on the same scale. A few similarities and a few significant 
differences can be noted. Generally both models show wavy structure - which is a well studied phenomenon 
\citep[see e.g.,][]{campo94,wyatt11} - but the exact structure of the waves differs. 

\begin{figure*}[!t]
\begin{center}
% ApJ Double Column scaling
\includegraphics[angle=0,scale=0.75]{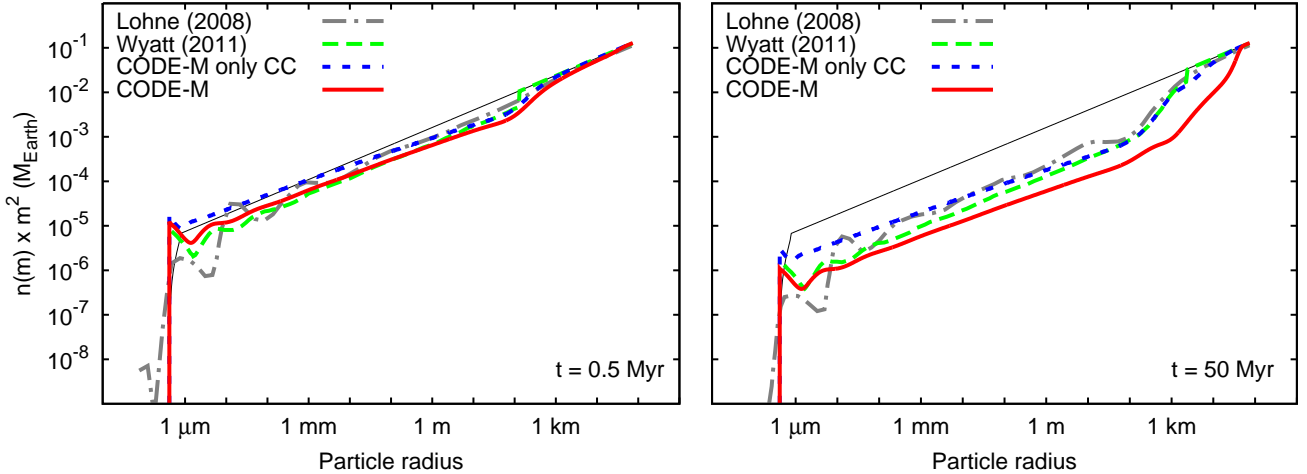}
% ApJ Manuscript scaling
%\includegraphics[angle=0,scale=0.7]{f8.eps}
\caption{Comparison of the evolution of the particle distribution within a debris disk 
around a solar-type star modeled by \cite{lohne08}, \cite{wyatt11} and the code 
presented in this paper. The "only CC" {\tt CODE-M} model uses only catastrophic collisions.
The thin solid line is the initial distribution. 
See text for more details.}
\label{fig:krivovm}
\end{center}
\end{figure*}

\begin{figure*}[!t]
\begin{center}
% ApJ Double Column scaling
\includegraphics[angle=0,scale=0.7]{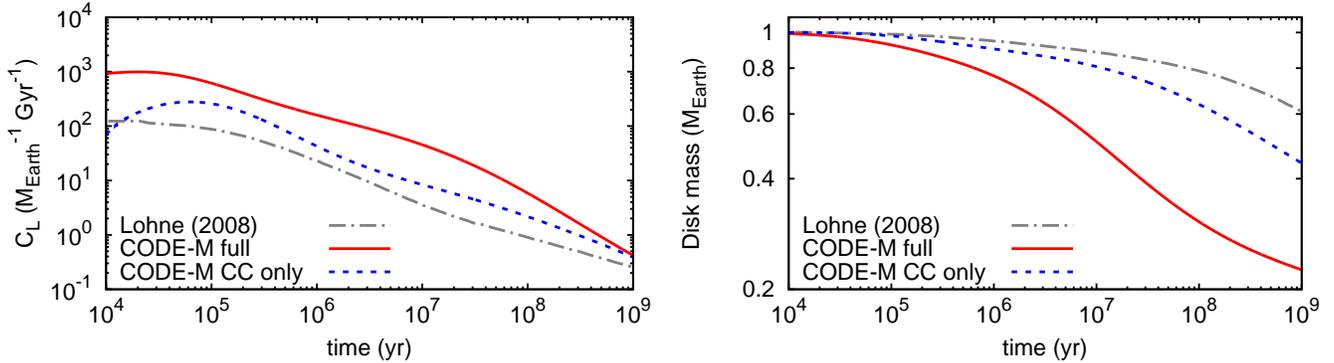}
% ApJ Manuscript scaling
%\includegraphics[angle=0,scale=0.65]{f9.eps}
\caption{Comparison of the evolution of the dust mass within a debris disk around a solar-type
star modeled by \cite{lohne08} and the model presented in this paper. See text for details.}
\label{fig:krivovmass}
\end{center}
\end{figure*}

Our modified erosive (cratering) prescription model gives a much better
agreement with the \cite{thebault03} results than our fiducial prescription, in the sense that it yields
a deeper first wave in the distribution with a larger wavelength. 
The offset between the locations of the first dip and the subsequent peak in the two models
could likely result from the higher collisional velocities that \cite{thebault03} calculate for the smallest
particles.
Our modified erosion prescription gives good agreement with the \cite{thebault03} results for particles
larger than a km in size, which is a surprise as the \cite{thebault03} erosive constants are for much softer materials
than our nominal values. Just above the blow-out regime our model becomes abundant in dust particles, as more and more dust is 
placed on highly eccentric orbits. Although some smoothing is expected in reality, we do expect the 
number of dust particles near the blowout limit to increase. 

\begin{figure}[!t]
\begin{center}
\includegraphics[angle=0,scale=0.68]{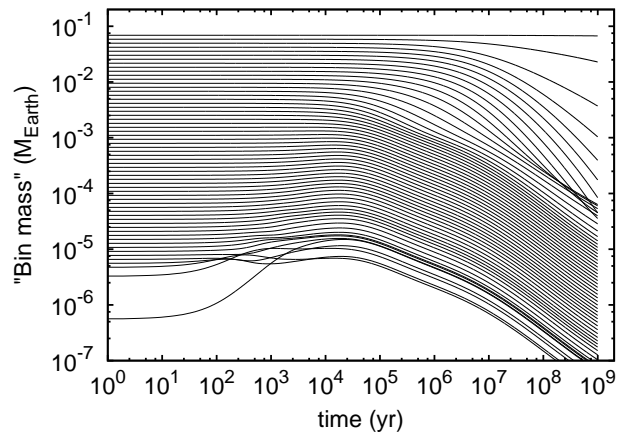}
\caption{Evolution of the total mass within consecutive mass regions from the smallest
to the largest particles in the system for the full collisional system, using the 
{\tt ii-0.3} parameters of \cite{lohne08}. The plot can be compared to the top panel 
of Figure 4.\ of \cite{lohne08}.}
\label{fig:bins}
\end{center}
\end{figure}

\begin{figure*}[!t]
\begin{center}
% ApJ Double Column scaling
\includegraphics[angle=0,scale=0.75]{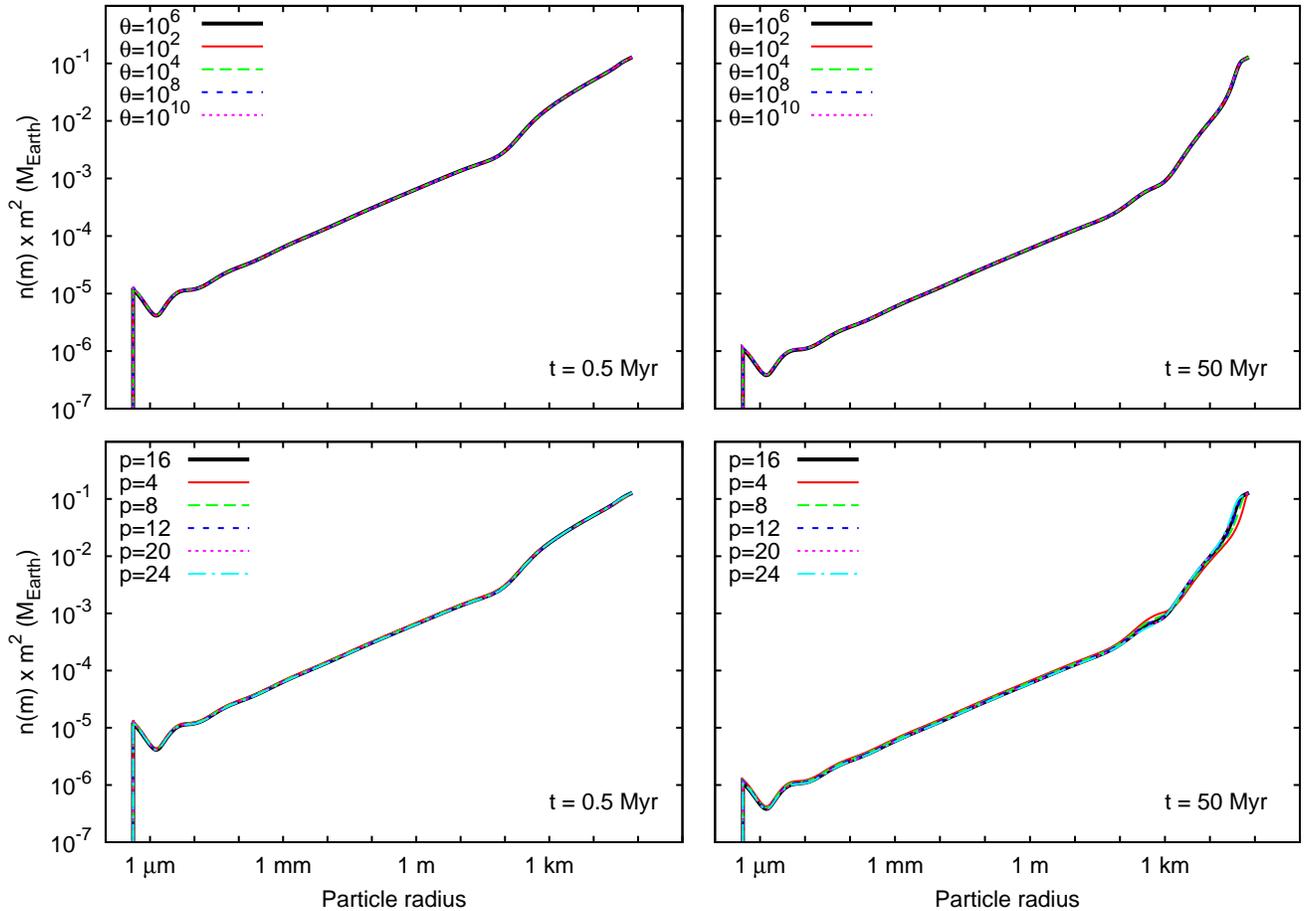}
% ApJ Manuscript scaling
%\includegraphics[angle=0,scale=0.7]{f11.eps}
\caption{Comparison of the evolution of our model distribution using the \cite{lohne08}
{\tt ii-0.3} parameters, when varying the parameters of the weights in the collision cross
sections for large particles.}
\label{fig:krivovm2}
\end{center}
\end{figure*}

While both our distributions show the typical double power-law feature of quasi steady-state 
collisional cascades \citep[see e.g.,][]{wyatt11} above and below the change in the 
strength curve, it is masked in the \cite{thebault03} model, due to the high amplitude wavy structures.
These structures are substantially reduced for the innermost ring of the disk modeled in \cite{thebault07},
but persist in the outer zones.

In Figure \ref{fig:thebaultmass}, we show the differences in the evolution of the disk mass between
the nominal case of \cite{thebault03} and our models. The left panel shows the evolution of the total
disk mass within the debris ring, while the right panel shows the evolution of the dust-to-planetesimal
mass ratio. These figures are equivalent to \cite{thebault03} figures 2 and 3 (except that
these are in plotted in logarithmic scales). Our nominal model predicts a faster decay of the total disk mass, reaching 25\% mass
loss, while the modified erosive prescription agrees with the Th\'ebault et al.\ (2003) model
and loses $\sim$ 12\% of its initial mass. The evolution of the 
dust-to-planetesimal disk mass differs while the quasi steady-state is being reached, after which all models decay with the same slope. 
Our nominal case model has an order of magnitude larger dust-to-planetesimal mass ratio at all times compared 
to the \cite{thebault03} model, while our modified erosive collision prescription case is close to it.

There are some easily identified differences between our models. \cite{thebault03} use the same
\cite{benz99} dispersive strength curve as we do, although they do average it to account for
impact angle variations. This is an unnecessary step, 
as the \cite{benz99} strength curve is already impact angle averaged, and is corrected in 
\cite{thebault07}. However, we find that this scaling offset does not have a 
significant effect on the outcome of the distribution evolution. \cite{thebault03} use a double 
power-law for fragment redistribution, while we use only a single power-law. 
We find that varying the slope of the single power-law does not have a significant effect on the 
evolution of the distribution either, so it appears that this difference is also likely 
not a significant contributing factor to the discrepancies.
A noteworthy difference between our models is that \cite{thebault03} calculate fragment re-accumulation,
while we do not. This is a possible explanation for our discrepancies at high masses,
and the offsets we have in the total mass decay.

The most significant difference between the models is that ours uses a single interaction
velocity, while \cite{thebault03} model the interaction velocity between the $\beta$ critical elliptical
orbit smallest particles and the parent ring. This is likely to account for some of the additional 
offsets for the smallest particles, as higher interaction velocities have been shown to initiate higher amplitude
waves \citep{campo94,wyatt11}. \cite{thebault03} also take into account the constant presence of 
particles smaller than the blowout limit within the parent ring, which we do not. 
Within denser disks this step may smooth out any features near the blowout limit.

\subsection{Comparison to \cite{lohne08} and \cite{wyatt11}}

\begin{figure*}[!t]
\begin{center}
% ApJ Double Column scaling
\includegraphics[angle=0,scale=0.7]{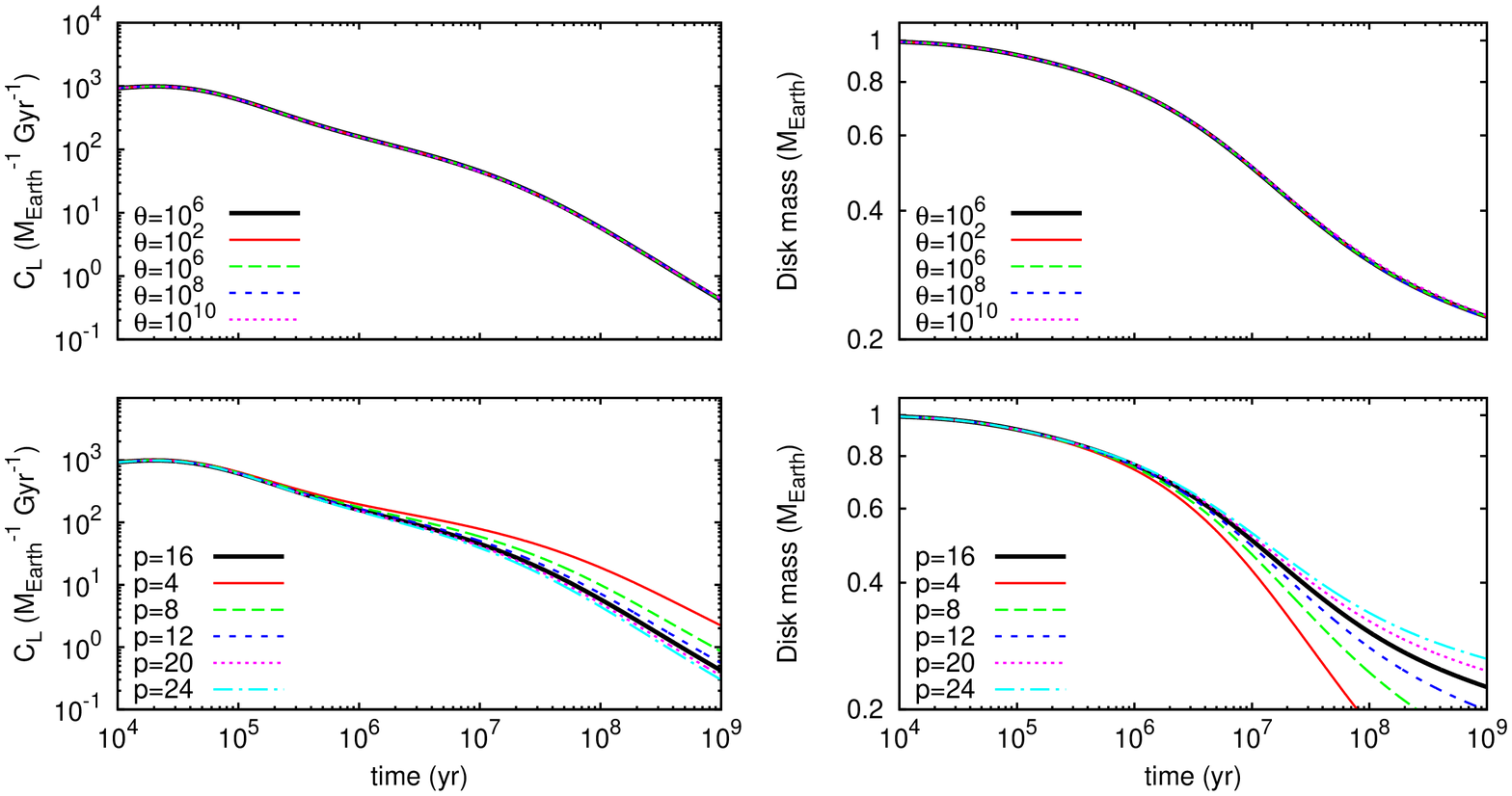}
% ApJ Manuscript scaling
%\includegraphics[angle=0,scale=0.65]{f12.eps}
\caption{The change in the dust mass evolution when varying the parameters of the weights in the collision cross
sections for large particles.}
\label{fig:krivovmass2}
\end{center}
\end{figure*}

\begin{figure}[!t]
\begin{center}
\includegraphics[angle=0,scale=0.7]{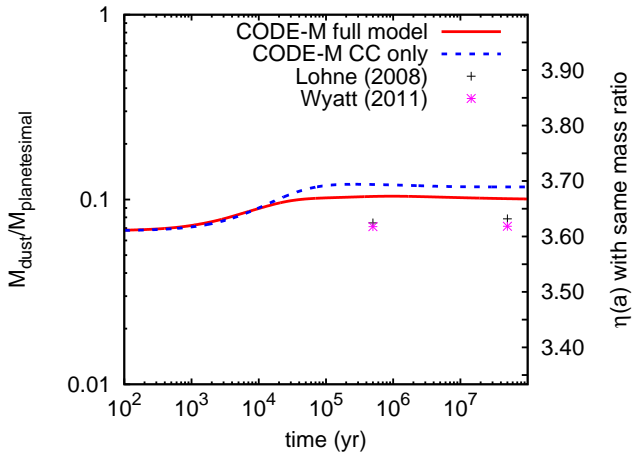}
\caption{The evolution of the dust-to-planetesimal mass ratio of the {\tt CODE-M} models and
the values for the \cite{lohne08} and \cite{wyatt11} models at 0.5 and 50 Myr. We calculate the
dust mass from 0.1 mm to 10 cm and the planetesimal mass from 10 cm to 100 m. On the right vertical
axis we give the value of the slope of the power-law distribution that would give the same dust-to-planetesimal
mass ratio.}
\label{fig:slopes}
\end{center}
\end{figure}
As introduced in \S 2, the numerical code {\tt ACE} 
\citep{krivov00,krivov05,krivov06,krivov08,lohne08,muller10} solves the dynamical evolution
of the collisional system as well as the collisional fragmentation, thus a straight
comparison to {\tt CODE-M} cannot be performed. We use their {\tt ii-0.3} model \citep{lohne08} for comparison, which is
for a relatively wide (7.5-15 AU), extremely dense (1 M$_{\earth}$ total mass with a largest
planetesimal size of 74 km) debris ring. We turned off the effects of Poynting-Robertson 
drag for this comparison model. The initial parameters we assumed are summarized in 
Table \ref{tab:comp}. This same system was modeled by \cite{wyatt11}, whose results
we also use for comparison.

In Figure \ref{fig:krivovm}, we show the evolution of the dust distribution of the system
given by {\tt CODE-M}, {\tt ACE}, and \cite{wyatt11}.
As the version of {\tt ACE} used in \cite{lohne08} only modeled catastrophic collisions and the \cite{wyatt11}
model uses catastrophic collision rates, we also include a {\tt CODE-M} run in the plot that only models
the outcomes of catastrophic collisions. Since the \cite{lohne08} values 
are already downscaled by ($\delta-1$), no additional scaling was required of
their data. The \cite{wyatt11} data points are divided by $\delta-1=0.0626$ to
convert them to differential number densities. Qualitatively, all distributions agree 
much better than in the \cite{thebault03} comparison (Figure \ref{fig:thebaultm}); however,
there is some scaling offset between the full 
{\tt CODE-M} and the other models, especially at large masses. 

The wavelengths of the waves roughly agree between {\tt CODE-M}
and {\tt ACE}, with the single difference being the absence of the strong offset of the
first crest in our model; the agreement is also good between {\tt CODE-M} and \cite{wyatt11}. 
The double power-law distribution due to the change from strength to gravity dominated 
thresholds in the strength curve \citep{benz99} can be distinguished
in all three models, with roughly the same slopes. The {\tt ACE} and the \cite{wyatt11} models
maintain their initial $-1.87$ number density distribution slope, 
while {\tt CODE-M} becomes somewhat steeper for the smallest particles.\footnote{In 
Figure \ref{fig:krivovm} we plot in the y-axis the product $n(m)m^2$, so that a steeper number density slope will show up as a flatter distribution.}
Our catastrophic-collision-only model has a smaller amplitude 
and wavelength wave structure than our full model or the other models.
The most significant difference between the models is the scaling offset of the full {\tt CODE-M} model, 
which we analyze below.

In Figure \ref{fig:krivovmass}, we show a comparison to figures 1 and 2 of \cite{lohne08}.
In the left panel we show the evolution of $C_L$ ($C_{\rm L{\ddot{o}}hne}$), which is 
introduced in Equation (11) of \cite{lohne08} as 
\begin{equation}
C_L = -\frac{\dot{M}_{\rm disk}}{M^2_{\rm disk}}\;,
\end{equation}
This quantity is inversely proportional 
to the characteristic timescale of the system. As expected, since our system evolves
faster, its characteristic timescale is shorter, so the $C_L$ factor for our models is larger.
This can be seen in the right panel as well, where we plot the decay of the total mass
in our system and that given in figure 2 of \cite{lohne08}. We adopted the
exact strength curve of \cite{lohne08} in this run of our model, with the
corrections given by \cite{wyatt11}.

In Figure \ref{fig:bins}, we show a comparison to the top panel of figure 4
in \cite{lohne08}, which shows the evolution of the total mass within each
of their mass bins. In this plot we show the evolution of the full collisional
system, which includes erosive and catastrophic collisions. Since we do not 
use mass bins, but rather a differential
number density griding, we integrate our distribution between 14 grid points
for each mass value, which roughly corresponds to a single mass bin of \cite{lohne08}.
Up to roughly a few hundred meters in size (where the strength curve has its minimum)
all mass ``bins" decay in close parallel slopes to each other after reaching
their quasi steady-state around 10,000 yr. This is in contrast to
the \cite{lohne08} results and agrees more with figure 2 of \cite{wyatt11},
who model the same system. Our intermediate size planetesimals ($\sim$ km) 
show a steeper decay than that modeled by either \cite{lohne08} or \cite{wyatt11}.

The obvious difference between {\tt ACE} and {\tt CODE-M}, is that {\tt ACE}
also evolves the dynamics in the system and takes into account the 
varying collisional velocities in the system from particles that are in
elliptical orbits within the parent ring. This could easily explain 
the offset of the first wave given by {\tt ACE} in Figure \ref{fig:krivovm}.

The increasing offset between the full {\tt CODE-M} run 
and the other two models is likely due to the omission of erosive collisions 
by \cite{lohne08} and to using catastrophic collision rates in \cite{wyatt11}. 
\cite{kobayashi10} have shown earlier erosive (cratering) collisions to be 
the dominant effect for mass loss in collisionally evolving systems. 
This effect is demonstrated by the {\tt CODE-M} model we run with only
catastrophic collisions included, which scales exactly with the {\tt ACE}
and \cite{wyatt11} models. Since {\tt CODE-M} does not include 
aggregation, the collisions of the smallest particles with the largest
bodies is not modeled perfectly. We assume the realistic distribution decay
to lie between the two models given by {\tt CODE-M}.

As introduced in \S \ref{largemass}, we artificially reduce the collision cross section of the largest particles in
the system to zero in order to avoid numerical instabilities.
However, as this is a completely arbitrarily defined numerical necessity, we investigate its
effects on the total mass decay, where we expect it to be the strongest. We reproduce
Figures \ref{fig:krivovm} \& \ref{fig:krivovmass2}, but with varying the values of $\Theta$ and $p$
in the smoothing function, in Figures \ref{fig:krivovm2} \& \ref{fig:krivovmass2}. 
As can be seen in these plots, the variable $\Theta$ does not affect either the evolution 
of the distribution or the total mass decay, as long as it is larger than one. Varying the values 
of $p$ does have an effect on both the evolution of the distribution and the total mass decay. The effect is only on
the largest bodies in the system; below a size of one hundred meters, the shapes of the distributions remain
unchanged, with the only differences being scaling offsets.

Visual examination of the distributions in Figure \ref{fig:krivovmass} hint at a
slightly steeper distribution slope for the {\tt CODE-M} models than the 
\cite{lohne08} and \cite{wyatt11} ones. Since the distributions have wavy structures
in them, this is difficult to show with a slope fit. Therefore, we calculate
the dust-to-planetesimal mass ratios of the distributions. We define the 
dust sizes to be from 0.1 mm to 10 cm and the planetesimal sizes to be from
10 cm to 100 m. In Figure \ref{fig:slopes} we plot the evolution of the mass
ratios of our models and the mass ratios for the \cite{lohne08} and \cite{wyatt11}
models at 0.5 and 50 Myr. Our models have significantly higher mass ratios
than theirs.  This is likely a result of the differences between our collisional equations.

\section{Conclusions}

In this paper we present a numerical model of the 
evolution of the distribution of dust in dense debris disks. 
We calculate our model with a new numerical code, {\tt CODE-M}, which we 
extensively verify and test in the Appendix.

Our collisional model and numerical solution both present improvements to
certain aspects of previous numerical collisional cascade models, but also 
have some limitations. Like most debris disk models \citep{thebault07,krivov06,lohne08,muller10,wyatt11},
we solve the Smoluchowski collisional equation \citep{smol16}, which does not enable the study of 
stochastic impact events. Our model also does not evolve the dynamical state of
the systems; the only code to currently do so is {\tt ACE} \citep{krivov05,lohne08,muller10}.
In its current state, our model is also limited to solving the collisional cascade in
debris rings, similarly to the initial versions of most collisional models \citep{krivov00,thebault03,wyatt11},
some of which have since been expanded to model extended disk structures \citep{krivov05,thebault07}.
Although our model only uses a single interaction velocity, as we show in \S \ref{sec:v},
this is a valid assumption as long as the interactions occur within the debris ring.

We introduce a new prescription for describing erosive collisions, which 
always takes into account the reduction of the mass in the largest fragment during 
the cratering event. We introduce a number of approximate interpolations
to ensure that our description of erosive and catastrophic collisions yields a continuous set of 
outcomes as a function of the colliding masses, while being consistent with experimental
results at various limits. Moreover, we employ an efficient numerical algorithm
that allows us to evaluate the scattering integrals with high precision, considering the
enormous dynamic range of masses involved in the calculation.

We compare our code to the previously published numerical models of collisional 
cascades in debris disks, reassuringly showing general agreement, particularly with 
\cite{lohne08} and \cite{wyatt11}. Nonetheless, our model shows faster decays 
than previously published ones \citep{thebault03,lohne08,wyatt11} and also yields slightly 
higher dust-to-planetesimal mass ratios. We attribute these characteristics to be a result of 
our accurate treatment of collisional cascades. Our 
model will be used in a series of upcoming papers to study those aspects of
debris disk behavior for which it is uniquely well suited.

\acknowledgments

We thank Richard Greenberg and David O'Brien for helpful discussions on collisional
outcomes and William Hartmann for advice on the smallest particles produced in collisions. 
We thank Philippe Th\'ebault, Alexander Krivov and Mark Wyatt for their comments on an earlier
version of the manuscript. Support for this work was provided by NASA through Contract Number 
1255094 issued by JPL/Caltech. 

\appendix

\section{Strength curves}

The redistribution outcome of collisions depends almost solely on the energy of the impact and
the colliding masses. In experiments it is common to specify the ratio of the kinetic energy of the projectile 
to the mass of the target. 
This ratio is known as the specific energy $Q_{\rm imp}$ of the impact. \cite{gault69} already noticed that the 
fragment distribution of particles depends on $Q_{\rm imp}$ (which they called ``rupture energy") when firing 
aluminum projectiles into glass targets. Their experiments showed that the fragments will have a power-law 
distribution, with the largest fragment being a function of the specific energy of the impact. This relationship
was first given in equation format in \cite{fujiwara77} for basalt targets. They note an offset from the \cite{gault69}
results, likely due to material strength differences.

Two specific values of $Q_{\rm imp}$ are used: $Q_S^{\ast}$ (the shattering specific
energy) and a somewhat larger $Q_D^{\ast}$ (the dispersion specific energy). The value of $Q_S^{\ast}$ gives
the energy required to shatter the target so that the mass of the largest fragment is no more than half of the original 
target mass. However, if the target is large enough, then self gravity pulls the fragments back together, 
leaving a remnant larger than half of the original. The larger $Q_D^{\ast}$ gives the value of $Q_{\rm imp}$ 
needed to disperse the fragments, so that the largest remaining piece is half of the original target mass. 
At lower target masses, where self-gravity can be neglected, $Q_D^{\ast} \approx Q_S^{\ast}$. We use 
$Q_D^{\ast}$ in our code and refer to it as $Q^{\ast}$.

Determining the value of $Q^{\ast}$ is difficult, especially for such a large range of particle sizes, from 
$\mu$m to km. The values for smaller bodies on the order of a few kilograms are mostly determined 
from laboratory experiments, while the values for larger bodies are determined from 
under-surface explosions, observations of large asteroids and with experiments done under very high pressure \citep{holsapple02}. 
However, material strength varies greatly as a function of material type, object size, surface type 
and the number of shattering events an object has gone through over its lifetime. An object that has gone through many
collisions in its lifetime, but still remains in one piece (descriptively called a ``rubble pile") can endure
harder collisions, which can actually be absorbed and help to compact the object, rather than dispersing it into 
smaller particles. This may seem like an important parameter only for larger objects; however, the evolution of 
larger objects significantly influences the evolution of smaller particles, and thus is important in our study.
We also lack experiments done with targets and impactors cooled down to space temperatures of 100-150 K,
where one would assume that objects get more brittle and easier to shatter.

Experiments clearly show that $Q^{\ast}$ is a function of the target mass $M$, meaning that different mass targets
will get shattered (with a $0.5 M$ largest fragment) by different specific energies. \cite{holsapple02} 
reviews experimental and theoretical results on collisions and strength curves. A common result for all of them 
is a minimum in the strength curve for bodies around 0.3 km in radius, where 
planetesimals are easiest to disperse (the number of cavities  and cracks weakening the bodies increases, while self-gravitation
is not dominant yet). As a result, there is a bump in the size distribution of minor planets in the solar 
system around this size. Smooth particle hydrodynamic (SPH) models give the $Q^{\ast}$ strength curve for 
larger bodies, while experiments help to anchor the curve for smaller rocks on the scale of a few cm in radius. 
It is still not clear whether the power-law shape of the 
curve can be extrapolated down to micron size particles, where experiments cannot be carried out. To study the 
collisional evolution of the smallest particles, the exact value of the strength curve must be known. In the absence 
of any models/experiments currently at those sizes, the best that can be done is a simple extrapolation of the strength 
curve to those regimes. \cite{stewart09} introduce a velocity-dependent tensile strength curve, that is defined by
variables such that it removes ambiguities over material density and projectile-to-target mass ratio. Their tensile
strength curve is ideal for low-velocity (1-300~m~s$^{-1}$) collisions, such as those found during planet
formation or at large radii debris disks. However, their universal relationship does not hold for conditions
that strongly depart from the catastrophic disruption regime ($Q_{\rm imp} < 0.1 Q^{\ast}$).

In our models we use the \cite{benz99} dispersion strength curve. It is derived from SPH models, represents a 
reasonable average of all previous strength curves, and is impact angle averaged. This curve can be written as 
(all units are in SI)
\begin{equation}
Q^{\ast}(a) = 10^{-4} Q_{\rm sc}\left[S \left(\frac{a}{1~{\rm cm}}\right)^{s} + G \rho \left(\frac
{a}{1~{\rm cm}}\right)^{g}\right] {\rm J~g}{\rm erg}^{-1}{\rm~kg}^{-1}\;,
\label{eq:BA99}
\end{equation}
where the fiducial values in the equation are given in Table \ref{tab:tabvar}.

\section{Mass conservation of the model}

A crucial test of any collisional code is for it to conserve the initial total mass of the system. Since particles are removed
at the low mass end, this behavior can be complicated to verify. However, a system can only maintain its total mass numerically, if its
collisional equations are mass conserving analytically. Here, we prove that our collisional equation is
mass conserving.

The collisional equation can be written as
\begin{eqnarray}
\frac{{\rm d}n(m)}{{\rm d}t} &=& - \int_0^{\infty} {\rm d}m^{\prime} n(m) n(m^{\prime}) \sigma(m,m^{\prime}) \nonumber \\
			     && + \int_0^{\infty} {\rm d}\mu \int_{\mu}^{\infty} {\rm d}M n(\mu)n(M)\sigma(\mu,M) \delta\left[X(\mu,M)-m\right] \nonumber \\
			     && + \int_0^{\infty} {\rm d}\mu \int_{\mu}^{\infty} {\rm d}M n(\mu)n(M)\sigma(\mu,M) R(m;\mu,M)\;,
\label{eq:start}
\end{eqnarray}
where $R(m;\mu,M)$ is the redistribution function to mass $m$ from $\mu + M$ collisions, such that
\begin{equation}
\int_0^{\infty} {\rm d}m R(m;\mu,M) m = \mu + M - X(\mu,M)\;,
\end{equation}
and $\delta$ is the Kronecker function. Multiplying Equation (\ref{eq:start}) by $m$ and integrating over d$m$ gives
\begin{eqnarray}
\frac{{\rm d}M}{{\rm d}t} = \frac{{\rm d}n(m)m}{{\rm d}t} &=& - \int_0^{\infty}{\rm d}m \int_0^{\infty} {\rm d}m^{\prime} n(m) n(m^{\prime}) \sigma(m,m^{\prime})m \nonumber \\
			     &&	+ \int_0^{\infty} {\rm d}\mu \int_{\mu}^{\infty} {\rm d}M n(\mu)n(M)\sigma(\mu,M)\int_0^{\infty}{\rm d}m \delta\left[X(\mu,M)-m\right]m \nonumber \\
			     && + \int_0^{\infty} {\rm d}\mu \int_{\mu}^{\infty} {\rm d}M n(\mu)n(M)\sigma(\mu,M)\int_0^{\infty}{\rm d}m R(m;\mu,M)m \;,
\end{eqnarray}
where 
\begin{equation}
\int_0^{\infty}{\rm d}m \delta\left[X(\mu,M)-m\right]m = X(\mu,M)\;,
\end{equation}
resulting in
\begin{equation}
\frac{{\rm d}M}{{\rm d}t} = - \int_0^{\infty}{\rm d}m \int_0^{\infty} {\rm d}m^{\prime} n(m) n(m^{\prime}) \sigma(m,m^{\prime}) m
			    + \int_0^{\infty} {\rm d} \mu \int_{\mu}^{\infty}{\rm d}M n(\mu)n(M)\sigma(\mu,M)(\mu+M)
\label{eq:start2}
\end{equation}
The first integral can be separated into two sections as
\begin{eqnarray}
\int_0^{\infty}{\rm d}m \int_0^{\infty} {\rm d}m^{\prime} n(m) n(m^{\prime}) \sigma(m,m^{\prime}) m &=&   \int_0^{\infty}{\rm d}m \int_0^{m} {\rm d}m^{\prime} n(m) n(m^{\prime}) \sigma(m,m^{\prime}) m \nonumber \\
												     && + \int_0^{\infty}{\rm d}m \int_m^{\infty} {\rm d}m^{\prime} n(m) n(m^{\prime}) \sigma(m,m^{\prime}) m\;.
\label{eq:sub}
\end{eqnarray}
Since $\sigma(\mu,M)$ is a symmetric function, we can swap the limits of integration for $m$ and $m^{\prime}$ in the second integral of 
Equation (\ref{eq:sub}) and, after making
a change of variables of $m=\mu$ and $m^{\prime}=M$ in the first and $m^{\prime}=\mu$ and $m=M$ in the second integral, the full equation becomes
\begin{equation}
\frac{{\rm d}M}{{\rm d}t} = \int_0^{\infty} {\rm d} \mu \int_{\mu}^{\infty}{\rm d}M n(\mu)n(M)\left[\sigma(\mu,M)\mu + \sigma(M,\mu)M-\sigma(\mu,M)(\mu+M)\right]\;.
\label{eq:end}
\end{equation}
Since the collisional cross section is completely symmetric, the integral itself becomes zero, thus proving that our equation is mass conserving.

\begin{figure}[!t]
\begin{center}
% ApJ Double Column scaling
\includegraphics[angle=0,scale=1.40]{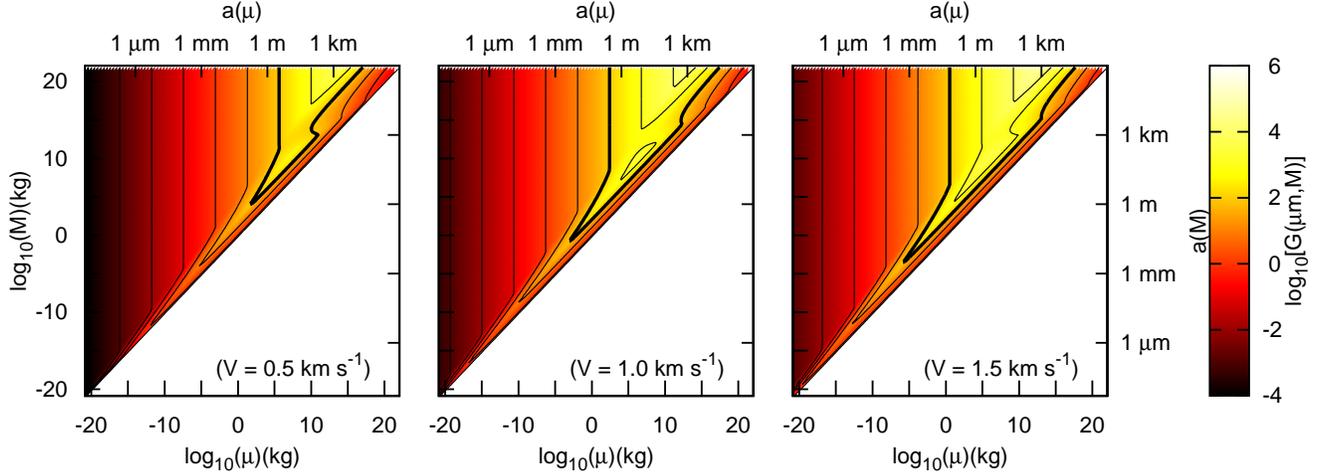}
% ApJ Manuscript scaling
%\includegraphics[angle=0,scale=1.30]{f14.eps}
\caption{The values of $G(m,m^{\prime})$ as a function of the colliding masses. The thick contour is for $G(m,m^{\prime})=100$, which is roughly equal to
the $\Gamma$ value used in \cite{dohnanyi69}. The panels give the contours as a function of collision velocities. 
The collisional velocities of 0.5, 1.0, and 1.5 km s$^{-1}$ correspond to debris ring radii of 100, 25, and 10 AU 
around an A spectral type star, respectively. The $G(m,m^{\prime})$ parameter is strongly dependent on the collisional velocity.}
\label{fig:Gplot}
\end{center}
\end{figure}

\section{Numerical evaluation of the model and verification tests}

The integro-differential equation presented in \S 3 must be integrated over 40 orders of magnitude in mass 
space, contains a double integral whose errors can easily increase if not evaluated carefully, and is bundled 
in a differential equation that evolves the number densities of dust grains and boulders within the same step.
These characteristics demand attention in its numerical evaluation. In the following subsections we explain the
numerical methods used to evaluate each integral and the ordinary differential equation (ODE). We also present 
verification and convergence tests for our code, which explain why such precisions are really necessary.

\subsection{Taylor series expansion of $T_{\rm I}$}

\begin{figure}[!t]
\begin{center}
% ApJ Double Column scaling
\includegraphics[angle=0,scale=1.40]{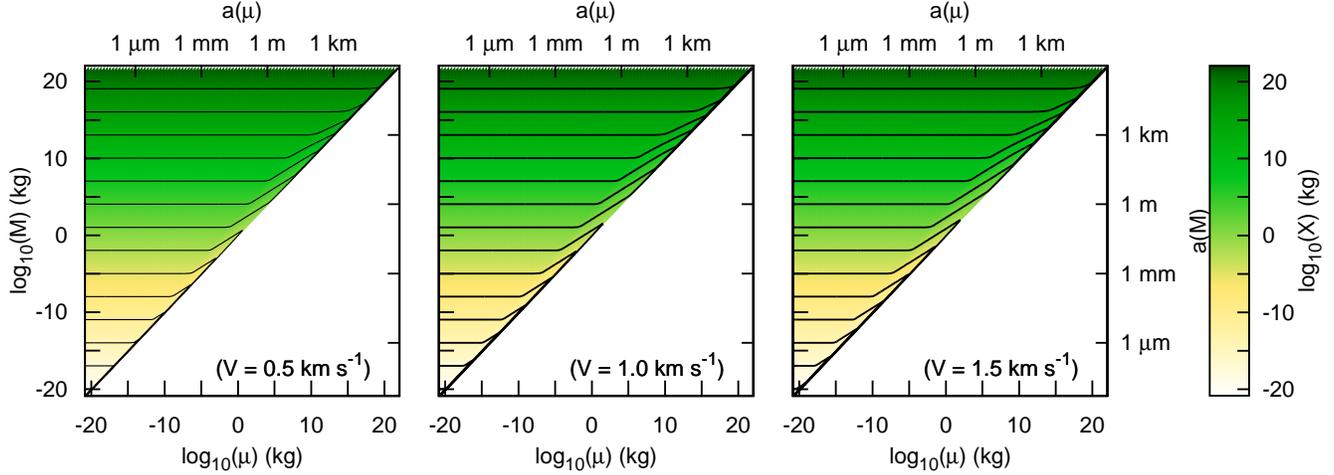}
% ApJ Manuscript scaling
%\includegraphics[angle=0,scale=1.30]{f15.eps}
\caption{The largest $X$ fragment produced by collisions between particles $\mu$ and $M$ as a function of collision velocities.
The collisional velocities of 0.5, 1.0, and 1.5 km s$^{-1}$ correspond to debris ring radii of 100, 25, and 10 AU 
around an A spectral type star, respectively.}
\label{fig:X}
\end{center}
\end{figure}

First, we expand equation (\ref{eq:t1}) to use a Taylor series when $m^{\prime} \ll m$ and $m^{\prime} < \mu_{\rm X}(m)$. 
For this, we rewrite $M$ in terms of $m$ and $m^{\prime}$ as
\begin{equation}
M=m+m^{\prime}G(m,m^{\prime})\;,
\end{equation}
where $m^{\prime}G(m,m^{\prime})$ equals the cratered mass, $m$ is the largest $X(M,m^{\prime})$ particle created and 
$G(m,m^{\prime})$ can be found by root finding algorithms. As written, $G(m,m^{\prime})$ can be related to the 
$\Gamma$ parameter used by \cite{dohnanyi69}, for which he used a constant value of 130 for $5~{\rm km}~{\rm s}^{-1}$ 
collisions. We plot the value of $G(m,m^{\prime})$ as a function of $\mu$ and $M$ in Figure \ref{fig:Gplot}, with the thicker solid line giving 
the contour of $G(m,m^{\prime})=100$. This contour lies at sizes reasonable for experiments in laboratory conditions, which is why 
\cite{dohnanyi69} used a value close to it. The positions of the contours are a strong function of the interaction velocities.
The $m^{\prime} < \mu_{\rm X}(m)$ integrand can be written as
\begin{eqnarray}
I(m,m^{\prime}) &=& f(m^{\prime}) w(m^{\prime}) \sigma_w(m^{\prime}){m^{\prime}}^{-\eta}a(m)^2\biggl(f(m)w(m)\sigma_w(m) \Bigl(1+Z\Bigr)^2\biggr. \nonumber \\
&& \biggl.-f(M) w(M) \sigma_w(M) \left[1+G(m,m^{\prime}) Z^3\right]^{-\eta}\Bigl\{Z+\left[1+G(m,m^{\prime})Z^3\right]^{\frac{1}{3}}\Bigr\}^2\biggr)
\end{eqnarray}
where $Z=a(m^{\prime})/a(m)$ and $f(m)$ and $f(m^{\prime})$ are dimensionless number densities that can be expressed as
\begin{equation}
f(m) = \frac{n(m)}{C m^{-\eta}}\;.
\end{equation}
We rewrite this integrand as
\begin{eqnarray}
I(m,m^{\prime}) &=& f(m^{\prime}) w(m^{\prime}) \sigma_w(m^{\prime}) {m^{\prime}}^{-\eta}a(m)^2f(m)w(m)\sigma_w(m)\Biggl(\left(1+Z\right)^2 \Biggr. \nonumber \\
&& \Biggl. -\frac{f(M)w(M)\sigma_w(M)}{f(m)w(m)\sigma_w(m)} \left[1+G(m,m^{\prime}) Z^3\right]^{-\eta}\left\{Z+\left[1+G(m,m^{\prime})Z^3\right]^{\frac{1}{3}}\right\}^2\Biggr)\;.
\end{eqnarray}
The Taylor series for the components are
%%%%%%%%%%%%%%%%%%%%%%%%%%%%%%%%%%%%%%%%%%%%%%%%%%%%
%
% Manuscript (Single Column) format of equation
%
%%%%%%%%%%%%%%%%%%%%%%%%%%%%%%%%%%%%%%%%%%%%%%%%%%
%\begin{eqnarray}
%\left(1+Z\right)^2 &=& 1 + 2Z + Z^2 \nonumber \\
%		   &\equiv& {\mathfrak T}_1
%\end{eqnarray}
%and
%\begin{eqnarray}
%\left[1+G(m,m^{\prime}) Z^3\right]{-\eta} \left[Z+\sqrt[3]{1+G(m,m^{\prime}) Z^3}\right]^2 &=& 1 + 2Z + Z^2 + \nonumber \\
%                                                                                             && \left[\frac{2G(m,m^{\prime})}{3} - \eta G(m,m^{\prime})\right]Z^3 \nonumber \\
%                    && + \left[\frac{2G(m,m^{\prime})}{3} - 2 \eta G(m,m^{\prime})\right] Z^4 \nonumber \\
%                    && - \eta G(m,m^{\prime}) Z^5 \nonumber \\
%		   &\equiv& {\mathfrak T}_1 + {\mathfrak T}_2			
%\end{eqnarray}
%%%%%%%%%%%%%%%%%%%%%%%%%%%%%%%%%%%%%%%%%%%%%%%%%%%%
%
% ApJ (Single Column Appendix) format of equation
%
%%%%%%%%%%%%%%%%%%%%%%%%%%%%%%%%%%%%%%%%%%%%%%%%%%%%
\begin{eqnarray}
\left(1+Z\right)^2 &=& 1 + 2Z + Z^2 \nonumber \\
		   &\equiv& {\mathfrak T}_1\\
\end{eqnarray}
and
\begin{eqnarray}
\left[1+G(m,m^{\prime}) Z^3\right]{-\eta} \left[Z+\sqrt[3]{1+G(m,m^{\prime}) Z^3}\right]^2 &=& 1 + 2Z + Z^2 + \left[\frac{2G(m,m^{\prime})}{3} - \eta G(m,m^{\prime})\right]Z^3 \nonumber \\
                    && + \left[\frac{2G(m,m^{\prime})}{3} - 2 \eta G(m,m^{\prime})\right] Z^4 - \eta G(m,m^{\prime}) Z^5 \nonumber \\
		   &\equiv& {\mathfrak T}_1 + {\mathfrak T}_2			
\end{eqnarray}
Both $f(M)/f(m)$ and $w(M)/w(m)$ are close to 1, while $\sigma_w(M)/\sigma_w(m)$ deviates from 1 as $m$ approaches $m_{\rm max}$. 
In those cases, the ratio can be expressed as
\begin{equation}
\frac{\sigma_w(M)}{\sigma_w(m)} = 1 + \left.\frac{\partial \sigma_w(m)}{\partial M}\right|_{M=m} (M-m) = 1 - P\frac{{\rm Exp}\left(-\frac{m_{\rm
max}-m}{\Theta}\right)}{\Theta\left[1-{\rm Exp}\left(-\frac{m_{\rm max}-m}{\Theta}\right)\right]} m^{\prime} G(m,m^{\prime})\;,
\end{equation}
since we know that $M-m=m^{\prime} G(m,m^{\prime})$. We write this ratio as
\begin{equation}
\frac{\sigma_w(M)}{\sigma_w(m)} = 1 - F\;.
\end{equation}
The integrand then takes the form
%%%%%%%%%%%%%%%%%%%%%%%%%%%%%%%%%%%%%%%%%%%%%%%%%%%%
%
% Manuscript (Single Column) format of equation
%
%%%%%%%%%%%%%%%%%%%%%%%%%%%%%%%%%%%%%%%%%%%%%%%%%%%%
%\begin{eqnarray}
%I(m,m^{\prime}) &=& f(m^{\prime}) w(m^{\prime}) \sigma_w(m^{\prime}){m^{\prime}}^{-\eta}a(m)^2 f(m) w(m) \sigma_w(m)\times \nonumber \\
%                && \biggl[{\mathfrak T}_1 - \left({\mathfrak T}_1 + {\mathfrak T}_2\right)\frac{w(M)f(M)}{w(m)f(m)}(1-F)\biggr]\;.
%\end{eqnarray}
%%%%%%%%%%%%%%%%%%%%%%%%%%%%%%%%%%%%%%%%%%%%%%%%%%%%
%
% ApJ (Single Column Appendix) format of equation
%
%%%%%%%%%%%%%%%%%%%%%%%%%%%%%%%%%%%%%%%%%%%%%%%%%%%%
\begin{equation}
I(m,m^{\prime}) = f(m^{\prime}) w(m^{\prime}) \sigma_w(m^{\prime}){m^{\prime}}^{-\eta}a(m)^2 f(m) w(m) \sigma_w(m)\times \biggl[{\mathfrak T}_1 - \left({\mathfrak T}_1 + {\mathfrak T}_2\right)\frac{w(M)f(M)}{w(m)f(m)}(1-F)\biggr]\;.
\end{equation}
Rearranging it gives us
%%%%%%%%%%%%%%%%%%%%%%%%%%%%%%%%%%%%%%%%%%%%%%%%%%%%
%
% Manuscript (Single Column) format of equation
%
%%%%%%%%%%%%%%%%%%%%%%%%%%%%%%%%%%%%%%%%%%%%%%%%%%%%
%\begin{eqnarray}
%I(m,m^{\prime}) &=& f(m^{\prime}) w(m^{\prime}) \sigma_w(m^{\prime}){m^{\prime}}^{-\eta}a(m)^2 f(m) w(m) \sigma_w(m)\times \nonumber \\
%                 && \Biggl\{{\mathfrak T}_1\left[1-\left(1-F\right)\frac{w(M)f(M)}{w(m)f(m)}\right] - {\mathfrak T}_2\frac{w(M)f(M)}{w(m)f(m)}(1-F)\Biggr\}
%\end{eqnarray}
%%%%%%%%%%%%%%%%%%%%%%%%%%%%%%%%%%%%%%%%%%%%%%%%%%%%
%
% ApJ (Single Column Appendix) format of equation
%
%%%%%%%%%%%%%%%%%%%%%%%%%%%%%%%%%%%%%%%%%%%%%%%%%%%%
\begin{equation}
I(m,m^{\prime}) = f(m^{\prime}) w(m^{\prime}) \sigma_w(m^{\prime}){m^{\prime}}^{-\eta}a(m)^2 f(m) w(m) \sigma_w(m)\times \Biggl\{{\mathfrak
T}_1\left[1-\left(1-F\right)\frac{w(M)f(M)}{w(m)f(m)}\right] - {\mathfrak T}_2\frac{w(M)f(M)}{w(m)f(m)}(1-F)\Biggr\}
\end{equation}
When
\begin{equation}
1-\frac{w(M)f(M)}{w(m)f(m)} < 10^{-9}\;,
\end{equation}
we use the approximate formula
\begin{equation}
I(m,m^{\prime})= f(m^{\prime}) w(m^{\prime}) \sigma_w(m^{\prime}){m^{\prime}}^{-\eta}a(m)^2 f(M) w(M) \sigma_w(m)\times \biggl[{\mathfrak T}_1 F - {\mathfrak T}_2(1-F)\biggr]\;.
\end{equation}
We use the Taylor series of the components to write the integrand below the limit of $Z < 10^{-3}$ (i.e., $m^{\prime}/m < 10^{-9}$). 
This means that our full integral for the first term ($T_{\rm I}$) takes the final form
\begin{equation}
\frac{{\rm d}f_I(m,t)}{{\rm d}t} = -V \pi {\rm C} \left\{ \int\limits_{m_{\rm min}}^{\mu_X(m)} {\rm d}m^{\prime} ~\mathcal{I} + \int\limits_{\mu_X(m)}^{m_{\rm max}} {\rm d}m^{\prime} f(m^{\prime},t)(m^{\prime})^{-\eta} f(m,t)\left(a(m)+a(m^{\prime})\right)^2\right\}
\label{eq:TI}
\end{equation}
where
%%%%%%%%%%%%%%%%%%%%%%%%%%%%%%%%%%%%%%%%%%%%%%%%%%%%
%
% Manuscript (Single Column) format of equation
%
%%%%%%%%%%%%%%%%%%%%%%%%%%%%%%%%%%%%%%%%%%%%%%%%%%%%
%\begin{equation}
%\mathcal{I} = \begin{cases}
%  f(m^{\prime}) w(m^{\prime}) \sigma_w(m^{\prime}){m^{\prime}}^{-\eta}a(m)^2 f(m) w(m) \sigma_w(m)\times \nonumber \\
%   \hspace{20mm} \Biggl\{{\mathfrak T}_1\left[1-\left(1-F\right)\frac{w(M)f(M)}{w(m)f(m)}\right] - {\mathfrak T}_2\frac{w(M)f(M)}{w(m)f(m)}(1-F)\Biggr\} \nonumber \\
%   \hspace{78mm}   \mbox{if } m^{\prime} < m\times10^{-9} \mbox{ \& } 1-\frac{w(M)f(M)}{w(m)f(m)} \ge 10^{-9} \\
%  f(m^{\prime}) w(m^{\prime}) \sigma_w(m^{\prime}){m^{\prime}}^{-\eta}a(m)^2 f(M) w(M) \sigma_w(m)\times \biggl[{\mathfrak T}_1 F - {\mathfrak T}_2(1-F)\biggr] \nonumber \\
%    \hspace{78mm}  \mbox{if } m^{\prime} < m\times10^{-9} \mbox{ \& } 1-\frac{w(M)f(M)}{w(m)f(m)} < 10^{-9} \\
%  f(m^{\prime},t)(m^{\prime})^{-\eta}\left[f(m,t)\left(a(m)+a(m^{\prime})\right)^2 - f(M,t)\left(\frac{M}{m}\right)^{-\eta}\left(a(M)+a(m^{\prime})\right)^2\right] \nonumber \\
%   \hspace{78mm}  \mbox{if } m\times10^{-9} \le m^{\prime} < \mu_X(m)
%\end{cases}
%\label{eq:T1I}
%\end{equation}
%%%%%%%%%%%%%%%%%%%%%%%%%%%%%%%%%%%%%%%%%%%%%%%%%%%%
%
% ApJ (Single Column Appendix) format of equation
%
%%%%%%%%%%%%%%%%%%%%%%%%%%%%%%%%%%%%%%%%%%%%%%%%%%%%
\begin{equation}
\mathcal{I} = \begin{cases}
  f(m^{\prime}) w(m^{\prime}) \sigma_w(m^{\prime}){m^{\prime}}^{-\eta}a(m)^2 f(m) w(m) \sigma_w(m)\times \nonumber \\
   \hspace{20mm} \Biggl\{{\mathfrak T}_1\left[1-\left(1-F\right)\frac{w(M)f(M)}{w(m)f(m)}\right] - {\mathfrak T}_2\frac{w(M)f(M)}{w(m)f(m)}(1-F)\Biggr\}
     & \mbox{if } m^{\prime} < m\times10^{-9} \mbox{ \& } 1-\frac{w(M)f(M)}{w(m)f(m)} \ge 10^{-9} \\
  f(m^{\prime}) w(m^{\prime}) \sigma_w(m^{\prime}){m^{\prime}}^{-\eta}a(m)^2 f(M) w(M) \sigma_w(m)\times \nonumber \\
    \hspace{20mm} \biggl[{\mathfrak T}_1 F - {\mathfrak T}_2(1-F)\biggr] & \mbox{if } m^{\prime} < m\times10^{-9} \mbox{ \& } 1-\frac{w(M)f(M)}{w(m)f(m)} < 10^{-9} \\
  f(m^{\prime},t)(m^{\prime})^{-\eta}\left[f(m,t)\left(a(m)+a(m^{\prime})\right)^2 - \right. \nonumber \\
  \hspace{20mm} \left. f(M,t)\left(\frac{M}{m}\right)^{-\eta}\left(a(M)+a(m^{\prime})\right)^2\right] & \mbox{if } m\times10^{-9} \le m^{\prime} < \mu_X(m)
\end{cases}
\label{eq:T1I}
\end{equation}
In Figure \ref{fig:X}, we show the mass of the $X$ fragments created when particles of mass $\mu$ and $M$ collide. 
The $m = X(\mu,M)$ regions are well defined in our single collisional velocity case. When using collisional 
velocity that depends on particle size, more than one $\mu_X(m)$ boundary may exist.

\subsection{Verification of the numerical precision of $T_{I}$}

To verify the precision of our integrator we set up an equation that is
similar to $T_{\rm I}$ in behavior and that has an analytic solution and compare
values given by our code to it.
The integral we evaluate both analytically and numerically with our code is
\begin{eqnarray}
TV_{\rm I}(m) 	&=& \int_{m_{\rm min}}^{\mu_{X(m)}}{\rm d}m^{\prime} (m^{\prime})^{-\eta}
\left\{\left(m^{\frac{1}{3}}+{m^{\prime}}^{\frac{1}{3}}\right)^2 - \left[(m+m^{\prime}\Gamma)^{\frac{1}{3}}+{m^{\prime}}^{\frac{1}{3}}\right]^2
\left(\frac{m+m^{\prime}\Gamma}{m}\right)^{-\eta}\right\} + \nonumber \\
		&& \int_{\mu_{X(m)}}^{m_{\rm max}}{\rm d}m^{\prime} (m^{\prime})^{-\eta} \left(m^{\frac{1}{3}}+{m^{\prime}}^{\frac{1}{3}}\right)^2 \;,
\end{eqnarray}
where we have removed all the constants and the dimensionless number densities.
We have also replaced $G(m^{\prime},m)$ with a constant $\Gamma$ 
(which can be related to the $\Gamma$ parameter used by Dohnanyi 1969) to enable
an analytic solution. To verify our algorithm for the evaluation of this term we
set the initial particle distribution to a power-law ($\eta$=11/6). The integration boundary can be evaluated as
\begin{equation}
\mu_X(m) = \begin{cases}
\frac{m}{1-\Gamma}      & \mbox{if } \Gamma < 1 \\
\frac{m_{\rm max} - m}{\Gamma}      & \mbox{if } \Gamma \ge 1
\end{cases}
\label{eq:mux_ver}
\end{equation}
The first integral is (setting $Z\equiv\frac{m^{\prime}}{m}$) 
%%%%%%%%%%%%%%%%%%%%%%%%%%%%%%%%%%%%%%%%%%%%%%%%%%%%
%
% Manuscript (Single Column) format of equation
%
%%%%%%%%%%%%%%%%%%%%%%%%%%%%%%%%%%%%%%%%%%%%%%%%%%%%
%\begin{eqnarray}
%{\mathfrak F} &=& -\frac{2}{5} Z^{-\frac{2}{3}} {m^{\prime}}^{-\frac{1}{6}} \Biggl\{ \left( 15 Z^{\frac{2}{3}} + 10 Z^{\frac{1}{3}} + 3 \right) - 
%    \left(1 +\Gamma Z \right)^{\frac{1}{6}} \times \Biggr. \nonumber \\
%    &&  \Biggl. \left[\frac{ 10 Z^{\frac{1}{3}}\left( 1 + 2 Z \Gamma\right)}{\left(1 + Z \Gamma\right)^{\frac{2}{3}}} 
%    + \frac{3\left( 1 + 6 Z \Gamma\right)}{\left( 1 + Z \Gamma\right)^{\frac{1}{3}}}
%    + \frac{3 Z^{\frac{2}{3}} \left( 5 + 6 Z \Gamma\right)}{ 1 + Z \Gamma}\right]\Biggr\}.
%\end{eqnarray}
%%%%%%%%%%%%%%%%%%%%%%%%%%%%%%%%%%%%%%%%%%%%%%%%%%%%
%
% ApJ (Single Column Appendix) format of equation
%
%%%%%%%%%%%%%%%%%%%%%%%%%%%%%%%%%%%%%%%%%%%%%%%%%%%%
\begin{equation}
{\mathfrak F} = -\frac{2}{5} Z^{-\frac{2}{3}} {m^{\prime}}^{-\frac{1}{6}} \Biggl\{ \left( 15 Z^{\frac{2}{3}} + 10 Z^{\frac{1}{3}} + 3 \right) - 
    \left(1 +\Gamma Z \right)^{\frac{1}{6}}\left[\frac{ 10 Z^{\frac{1}{3}}\left( 1 + 2 Z \Gamma\right)}{\left(1 + Z \Gamma\right)^{\frac{2}{3}}} 
    + \frac{3\left( 1 + 6 Z \Gamma\right)}{\left( 1 + Z \Gamma\right)^{\frac{1}{3}}}
    + \frac{3 Z^{\frac{2}{3}} \left( 5 + 6 Z \Gamma\right)}{ 1 + Z \Gamma}\right]\Biggr\}.
\end{equation}
When $\frac{m^{\prime}}{m} = Z < 10^{-9}$, the equation does not describe the analytic result correctly as the
first three components completely cancel the last three components numerically; however, analytically the result is 
not zero and this non-zero component gets multiplied by a large number. This is the same catastrophic cancellation that
affects the numerical evaluation of $T_{\rm I}(m)$. To overcome this and correctly represent the analytic result of the integral in
such cases, we rewrote this to a Taylor series as well. The first three components cancel, and we are left with
\begin{equation}
{\mathfrak F} = +\frac{2}{5} Z^{-2/3} {m^{\prime}}^{-1/6}\Biggl[\frac{35}{2}\Gamma Z + 15 \Gamma Z^{4/3} + \frac{11}{2} \Gamma Z^{5/3} 
  -\frac{65}{24}\Gamma^2Z^2 -\frac{25}{4}\Gamma^2 Z^{7/3} - \frac{85}{24}\Gamma^2 Z^{8/3} + \frac{665}{432}\Gamma^3Z^3 \Biggr]\;.
\end{equation}
The second integrand has a much simpler anti-derivative
\begin{equation}
{\mathfrak F} = -\frac{6}{5}\frac{m^{2/3}}{{m^{\prime}}^{5/6}} -4\frac{m^{1/3}}{{m^{\prime}}^{1/2}} - \frac{6}{{m^{\prime}}^{1/6}}
\end{equation}

\begin{figure}[!t]
\begin{centering}
% ApJ Double Column scaling
\includegraphics[angle=0,scale=0.73]{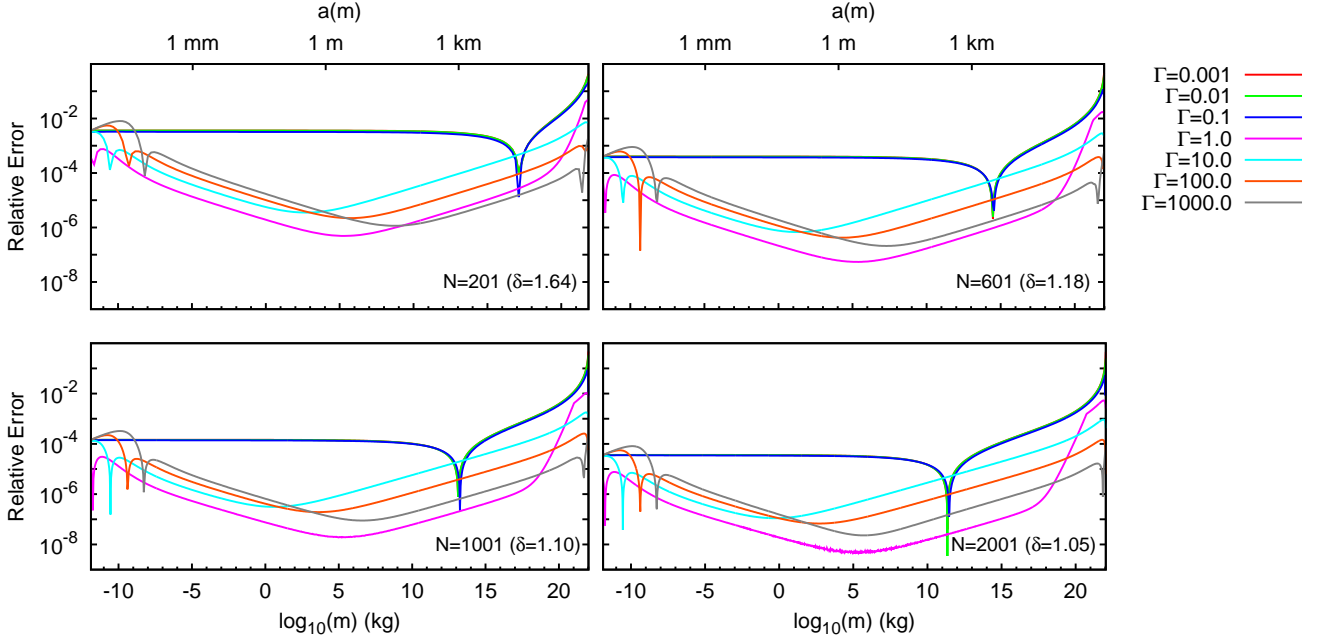}
% ApJ Manuscript scaling
%\includegraphics[angle=0,scale=0.65]{f16.eps}
\caption{The error in the integration of $T_{\rm I}$ as a function of the mass $m$ and the value
of the $\Gamma$ constant for neighboring mass grid ratios of $\delta=$1.64, 1.18, 1.10 and 1.05.}
\label{fig:T1eps}
\end{centering}
\end{figure}

In Figure \ref{fig:T1eps}, we show the computational error as a function of mass $m$, the $\Gamma$ constant and the number of grid 
points used (neighboring grid point mass ratio). 
In the actual model $\Gamma$ is not a constant, but equal to the variable $G(m,m^{\prime})$ which, as shown in 
Figure \ref{fig:Gplot}, varies from $10^{-4}$ to $10^{6}$. Figure \ref{fig:T1eps} shows that the errors do not
improve much past N=1000 ($\delta = 1.1$) and that, in general, errors are smaller for large values of 
$\Gamma$. This shows that errors originating from $T_{\rm I}$ are most likely to affect the smallest
particles in the system. We expect the errors to actually be completely symmetric, with the highest masses showing the
same quantitative errors as the lowest masses near the boundary. Offsets are due to the fact that our analytic model
includes targets larger than $m_{\rm max}$. The maximum error of $10^{-4}$ is not ideal, but acceptable.
When running our code, we use N=1000 grid points, which corresponds to a $\delta=1.1$.

\subsection{Numerical evaluation of $T_{II}$}

\begin{figure*}[!t]
\begin{center}
% ApJ Double Column scaling
\includegraphics[angle=0,scale=1.40]{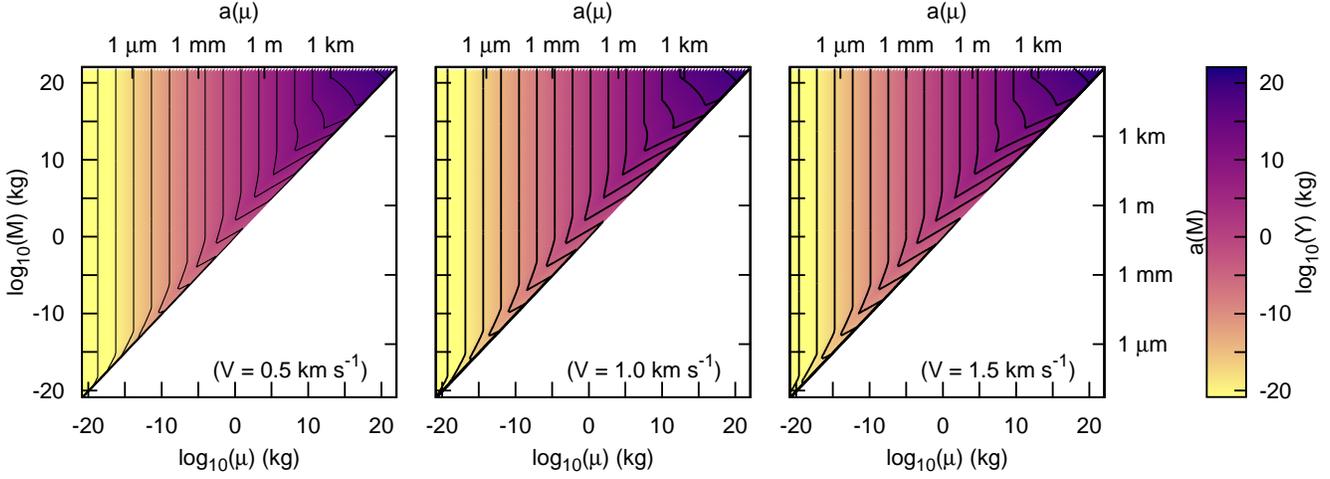}
% ApJ Manuscript scaling
%\includegraphics[angle=0,scale=1.30]{f17.eps}
\caption{Iso-size contours for the produced $Y$ fragments as a function of the colliding body sizes and interaction velocities. 
Fragments of size a(m) will be produced within regions where $a(m) < a(Y)$. The largest fragments produced are not heavily dependent on the 
interaction velocities. The collisional velocities of 0.5, 1.0 and 1.5 km s$^{-1}$ correspond to debris ring radii of 100, 25 and 10 AU 
around an A spectral type star, respectively.}
\label{fig:Yregions}
\end{center}
\end{figure*}

As a double integral, the second term, $T_{II}$, poses a larger challenge to achieve
acceptable precision. A collision between masses $\mu$ and $M$ will be able to
produce a mass $m$ in the redistribution power-law, if $m<Y(\mu,M)$. In Figure 
\ref{fig:Yregions} we plot the iso-$Y$ contours in the $\mu$ vs.\ $M$ phase space,
which shows that integrating between exact boundaries is difficult for $T_{II}$,
especially if the collisional velocity is not a constant but a function of the
particle  mass.

\begin{figure}[!ht]
\begin{centering}
\includegraphics[angle=0,scale=0.4]{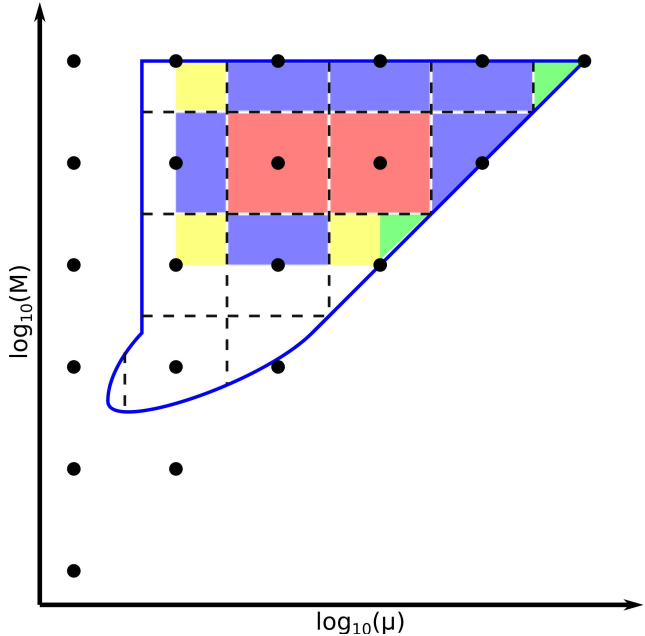}
\caption{Description of the integration method used for T$_{\rm II}$. The blue line represents the boundary, within which collisions 
are able to produce a certain mass $m$ in the redistribution power-law. Resolution elements capable of producing $m$ on their full 
area are colored red. Boundary resolution elements (i.e., $\mu \equiv M$ or $M \equiv m_{\rm max}$) will be able to produce $m$ on 
a ``half" area, colored blue. The tip of the distribution (green) will be able to produce on an eighth of its full area. Partial 
quarter contributions are given by the yellow areas. Increasing the number of grid points used obviously increases not just the 
precision but the area used for the integration as well.}
\label{fig:T2exp}
\end{centering}
\end{figure}

As a first step, we determine which $m$ masses can be produced by the grid points ($\mu$, $M$) and their
neighbors. For a grid point to be able to produce a particle of mass $m$, $Y(\mu,M)$ has to be larger than $m$. We determine the
limiting mass that is produced by each grid point and all of their neighbors as well. Up to that min$(\mu,M)$
value, all $m$ masses are produced with the full weight of the grid point. Between min$(\mu,M)$ and max$(\mu,M)$
- which is the largest $m$ still produced by the $(\mu,M)$ grid point itself - we analyze the areas divided
into quadrants, and assign integration weights appropriately. A simple plot is shown in Figure \ref{fig:T2exp} 
to explain the weights given to each grid point.

These minimum and maximum masses are usually at most 2-3 grid points 
apart. This means that on average 2-3 numbers have to be stored for all ($\mu$; $M$) grid points, as
below min$(\mu,M)$ all $m$ masses have the same weight. 
The final integration speed can be increased by factors of 5 as 
the integration loops can be run in non-redundant ways.

\subsection{Convergence tests}

\begin{figure}[!ht]
\begin{centering}
% ApJ Double Column scaling
\includegraphics[angle=0,scale=0.7]{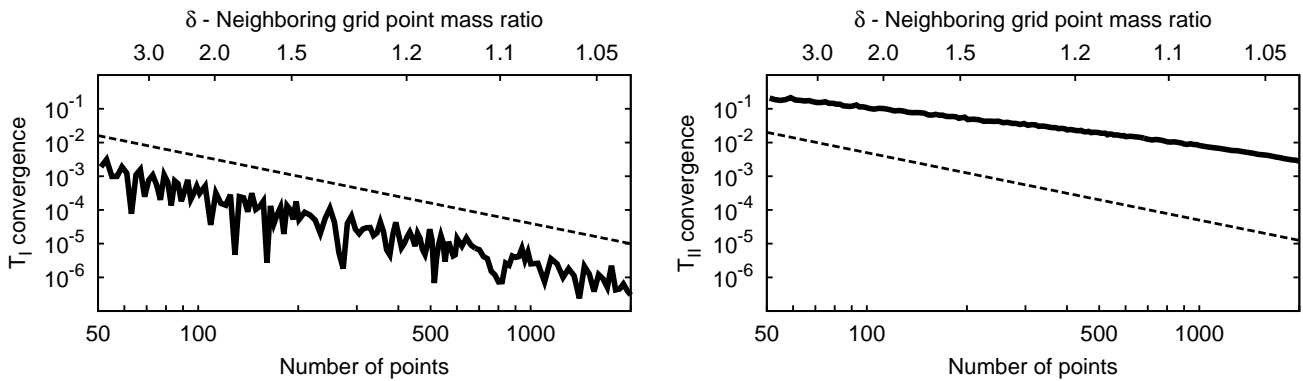}
% ApJ Manuscript scaling
%\includegraphics[angle=0,scale=0.65]{f19.eps}
\caption{Convergence test results of our code. The {\it left panel} gives the results for $T_{\rm I}$, while
the {\it right panel} for $T_{II}$. We also plot a $N^{-2}$ curve with a dashed line, which is the 
effective accuracy of the trapezoid integration method. The accuracy of the first term follows this trend, however, that of the second
term is shallower, due to the resolution dependent integration limits.}
\label{fig:Tconv}
\end{centering}
\end{figure}

We run convergence tests on our code for both terms to find the least
number of grid points we can use and still keep an acceptable numerical accuracy. We calculate
convergence to the value given using 4000 grid points. Our convergence plot for $T_{\rm I}$
in Figure~\ref{fig:Tconv} shows that, for such a large dynamic range in masses, one needs 
a neighboring grid mass ratio of at most $\delta=1.1$ to reach a relative error of $10^{-5}$.

Our convergence test for $T_{II}$ does not reach the same level of accuracy, as at $\delta \approx 1.1$ we reach
a relative error of $10^{-2}$ only. However, this error is driven by the resolution dependent
integration limits and not the method itself. As such the number of particles added by
$T_{II}$ will always be underestimated by a small amount.

\subsection{The ODE solver}

Previous work \citep[i.e.,][]{thebault03,thebault07,krivov00,lohne08}
used only a first order Eulerian algorithm to solve the differential
equation. We are using a 4$^{\rm th}$ order Runge-Kutta algorithm (RK4). 
To verify the ODE solver we simply evolve the Poynting-Robertson drag term, 
whose analytic solution is
\begin{equation}
n(m,t)=n(m,t=0)\exp\left(-\frac{t}{\tau_{\rm PRD}}\right)\;.
\end{equation}
In the following, we verified the accuracy of the ODE integrator by
setting $\beta \equiv 0.100$ for the particles and using the solar system
timescale of 400 years. 

\begin{figure}[ht]
\begin{center}
% ApJ Double Column scaling
\includegraphics[angle=0,scale=0.7]{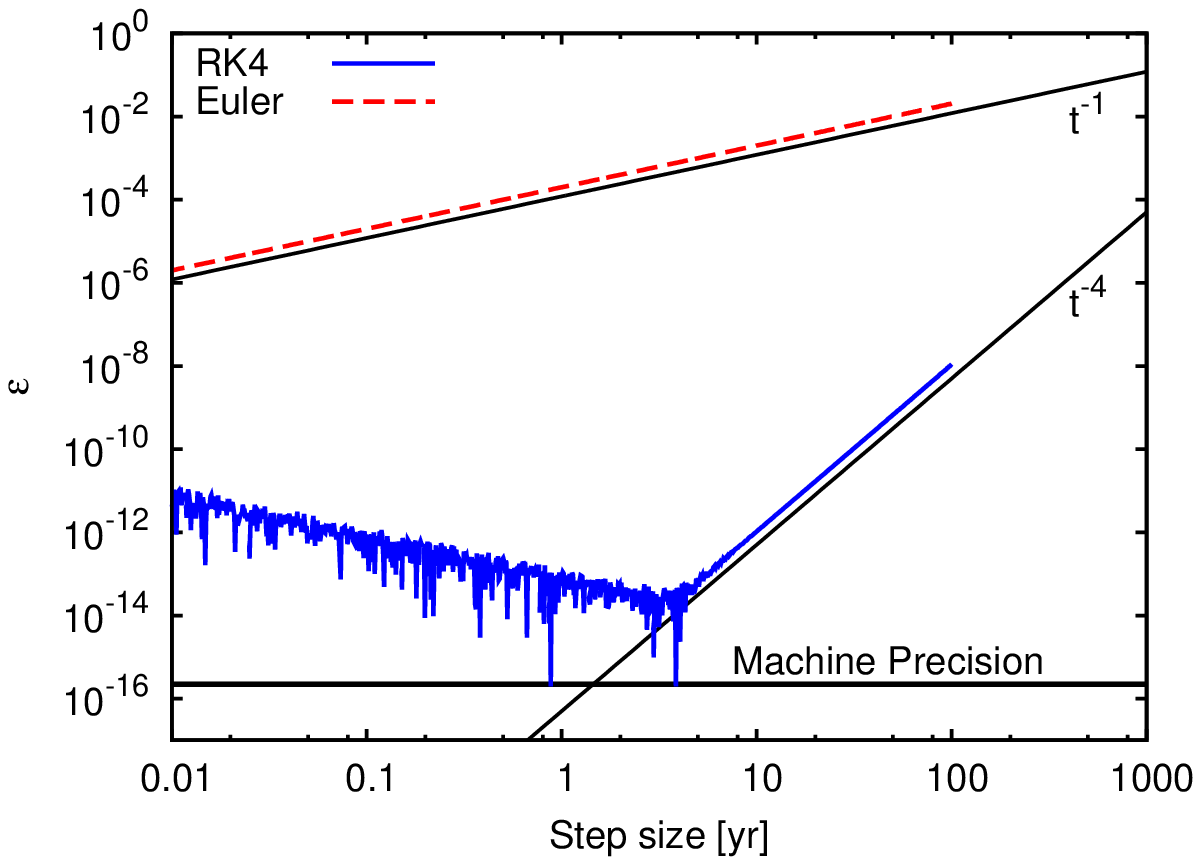}
\includegraphics[angle=0,scale=0.7]{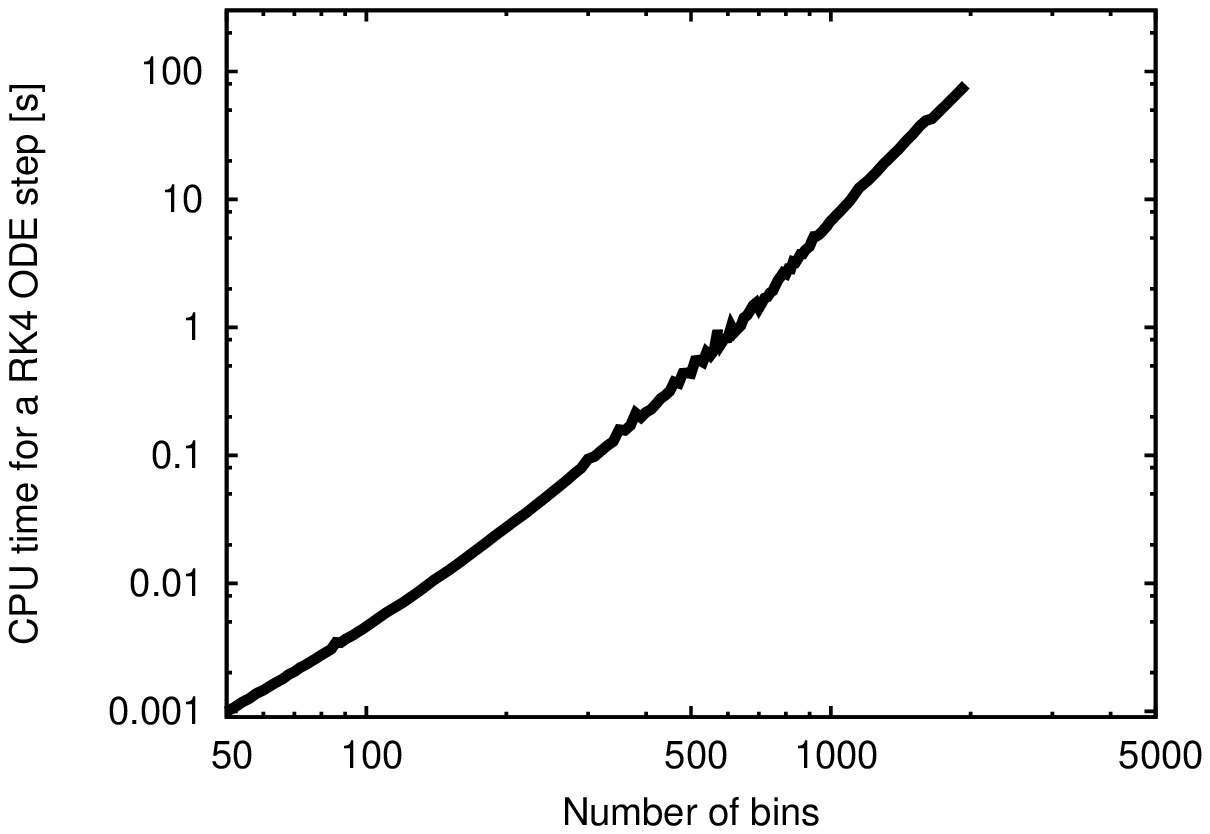}
% ApJ Manuscript scaling
%\includegraphics[angle=0,scale=0.6]{f20a.eps}
%\includegraphics[angle=0,scale=0.6]{f20b.eps}
\caption{{\it Left Panel:} The fractional difference between the numerical and analytical
results at a time roughly equal to 4000 yr as a function of the
time step used. We show errors for both an Euler ODE solver and for our
RK4 algorithm. We also plot a $t^{-1}$ and a $t^{-4}$ curve to guide the eye.
{\it Right Panel:} The amount of processor time needed by our code
to complete an RK4 step as a function of the number of mass grid points used.}
\label{fig:rk4}
\end{center}
\end{figure}

Using the results from the code, we define the ratio of the particle
density at some time $t$ to the particle density at time zero, i.e.,
$R_{\rm code}\equiv f(m,t)/f(m,t=0)$, and compare it to the analytic
result, $R_{\rm exp}=\exp(-t/\tau_{\rm PRD})$. We then compute the
fractional error between the numerical and the analytic solution.
The left panel of Figure \ref{fig:rk4} shows the fractional error as a function of 
the time step $\Delta t$, evaluated at a time roughly 
$t\approx\tau_{PRD}$ (4000 years) for both the RK4 and Euler method.

For this particular set-up, the optimal time step is $\sim
4$~yr for the RK4. However, the optimal time step depends on the time $t$ at
which the fractional difference is evaluated because of the
accumulation of round-off errors. 
Finally, the right panel of Figure \ref{fig:rk4} shows the amount of CPU time taken
to calculate an RK4 step as a function of the number of mass grid points used on a
Mac Pro4 with 2 2.26 GHz Quad Core Intel Xeon processors.

\end{document}